# AI-assisted inverse design of sequence-ordered high intrinsic thermal conductivity polymers

Xiang Huang[1], C. Y. Zhao[1], Hong Wang[2], and Shenghong Ju[1, 2, *]

[1] China-UK Low Carbon College, Shanghai Jiao Tong University, Shanghai, 201306, China

[2] Materials Genome Initiative Center, School of Material Science and Engineering, Shanghai Jiao Tong University, Shanghai, 201306, China

* Corresponding email: shenghong.ju@sjtu.edu.cn

## Abstract

Artificial intelligence (AI) promotes the polymer design paradigm from a traditional trial-and-error approach to a data-driven style. Achieving high thermal conductivity (TC) for intrinsic polymers is urgent because of their importance in the thermal management of many industrial applications such as microelectronic devices and integrated circuits. In this work, we have proposed a robust AI-assisted workflow for the inverse design of high TC polymers. By using 1144 polymers with known computational TCs, we construct a surrogate deep neural network model for TC prediction and extract a polymer-unit library with 32 sequences. Two state-of-the-art multi-objective optimization algorithms of unified non-dominated sorting genetic algorithm III (U-NSGA-III) and q-noisy expected hypervolume improvement (qNEHVI) are employed for sequence-ordered polymer design with both high TC and synthetic possibility. For triblock polymer design, the result indicates that qNHEVI is capable of exploring a diversity of optimal polymers at the Pareto front, but the uncertainty in Quasi-Monte Carlo sampling makes the trials costly. The performance of U-NSGA-III is affected by the initial random structures and usually falls into a locally optimal solution, but it takes fewer attempts with lower costs. 20 parallel U-NSGA-III runs are conducted to design the pentablock polymers with high TC, and half of the candidates among 1921 generated polymers achieve the targets (TC > 0.4 W m$^{-1}$K$^{-1}$ and SA < 3.0). Ultimately, we check the TC of 50 promising polymers through molecular dynamics simulations and reveal the intrinsic connections between microstructures and TCs. Our developed AI-assisted inverse design approach for polymers is flexible and universal, and can be extended to the design of polymers with other target properties.

# 1 Introduction

The compositional and structural diversity of polymers allows for highly tunable physical and chemical properties, and have a wide application in our daily lives [1-3]. Simultaneously, the near-diffuse chemical space of polymers makes it challenging to achieve specific properties in reality. The conventional Edisonian trial-and-error approach fails to match the urgent demands of the advanced polymer industry, since it is a long-term, costly and uncertain process [4,5]. Data-driven technology equipped with artificial intelligence (AI) as a powerful engine has been successfully utilized in the efficient development of polymers with desired properties [6-12]. Applying machine learning (ML) to the polymer community is mainly categorized into "forward problems" of high-throughput screening cases [13-17] and " inverse problems" of goal-oriented active design cases [18-23]. The basis for performing high-throughput screening of ideal polymers is the creation of a high-fidelity ML surrogate predictive model, which is trained on a certain amount of well-labelled data. However, these have been limited by the explorable chemical space, as the polymer candidates are manually collected [24]. If the polymer is not contained in the predefined library, the search is impossible to find it [25].

Another more universal and appropriate strategy is inverse design, where the desired property level is set in advance and the goal is achieved by combining polymer generation algorithms with optimization iterations. Recently, deep generative models, such as variational autoencoders (VAE) [26], recurrent neural networks (RNN) [27] and generative adversarial networks (GAN) [28], have been successfully applied in polymer research. However, the training of these models still requires a large amount of polymer structural data and their application in macromolecular design is still in the infancy stage. Due to the tight linkage between the structure and properties of polymers, it is possible to achieve performance enhancement by reorganization of some promising polymer sequences. This efficient and lightweight sequence-controlled technology has been successfully extended to the optimization of polymers with various superior properties, such as refractive index [29], bandgap [30] and glass transition temperature [31].

Thermal conductivity (TC) is one of the fundamental properties of polymers. The TC of intrinsic polymers is usually considered to be thermally insulating (less than 0.40 W m$^{-1}$K$^{-1}$) and therefore has been neglected for a long time in the past [32-34]. Yet, achieving high TC in polymers is urgently desired for fields such as organic electronics heat dissipation [35] and integrated circuit packaging [36]. Some efforts have been made to achieve the active design of polymers with high TC using ML in recent years [37-39]. Zhou et al. [37] employed genetic algorithms and molecular dynamics to design high TC polyethylene−polypropylene (PE-PP) copolymers. The TC of optimal sequence obtained at the 20th generation is 0.104 W m$^{-1}$K$^{-1}$, which was enhanced by about 700% and 45% compared with PE and PP homopolymers, respectively. Ma et al. [38] combined RNN and reinforcement learning to develop high TC polymers, and the best candidate has a molecular dynamics (MD) calculated TC of 0.69 W m$^{-1}$K$^{-1}$. Nagoya et al. [39] applied the Mont Carlo tree search algorithm to optimize the sequence of polyimide fragments. After about 1000 MD evaluations resulting in the best TC of 0.25 W m$^{-1}$K$^{-1}$. Despite these advances being valuable in guiding the development of high TC polymers, we believe that more efforts are required to enrich the dataset of polymers with high TC.

Herein, we have proposed and developed a AI-assisted workflow combining polymer fragment extraction, active optimization algorithms and molecular dynamics simulations for the inverse design of promising polymers with high TC outlined in Fig. 1. Our work starts from 1144 polymer data with MD-calculated TC in a recently publicized computational database [40]. Considering the costly polymer TC calculations, we first trained a deep neural network (DNN) agent model for simulating the TC of the emerging polymers in place of MD simulations using these data and Morgan fingerprints with frequency (MFF) [5]. The MFF captures the chemical substructures that appear in repeating units, and their contribution to the promotion/ inhibition of TC was analyzed by DNN with shapley additive explanations (SHAP) [41]. Combining the SHAP outputs and structural features of high TC polymers (TC ≥ 0.40 W m$^{-1}$K$^{-1}$ ) from the 1144 polymers, we constructed a polymer-unit library, including 32 potential small fragments, and binary coded them based on the serial number from [00000] to [11111] (see Fig. 1a). We then built two multi-objective optimization algorithms in Fig. 1b, the multi-objective evolutionary algorithm (MOEA) and multi-objective Bayesian optimization (MOBO), as we not only consider the TC but additionally evaluate the synthesizability of the new polymers. The synthesizability of polymers was evaluated by the SA score, which is based on molecular complexity and fragment contributions [42]. We measured the performance of the two algorithms on a complete triblock polymer dataset and further extended MOEA to a pentablock polymer design with more than tens of millions of possible sequences. Ultimately, we employed nonequilibrium molecular dynamics (NEMD) simulations to calculate the TCs of a batch of promising polymers and provide insights into the microscopic associations between TC and chain conformation.

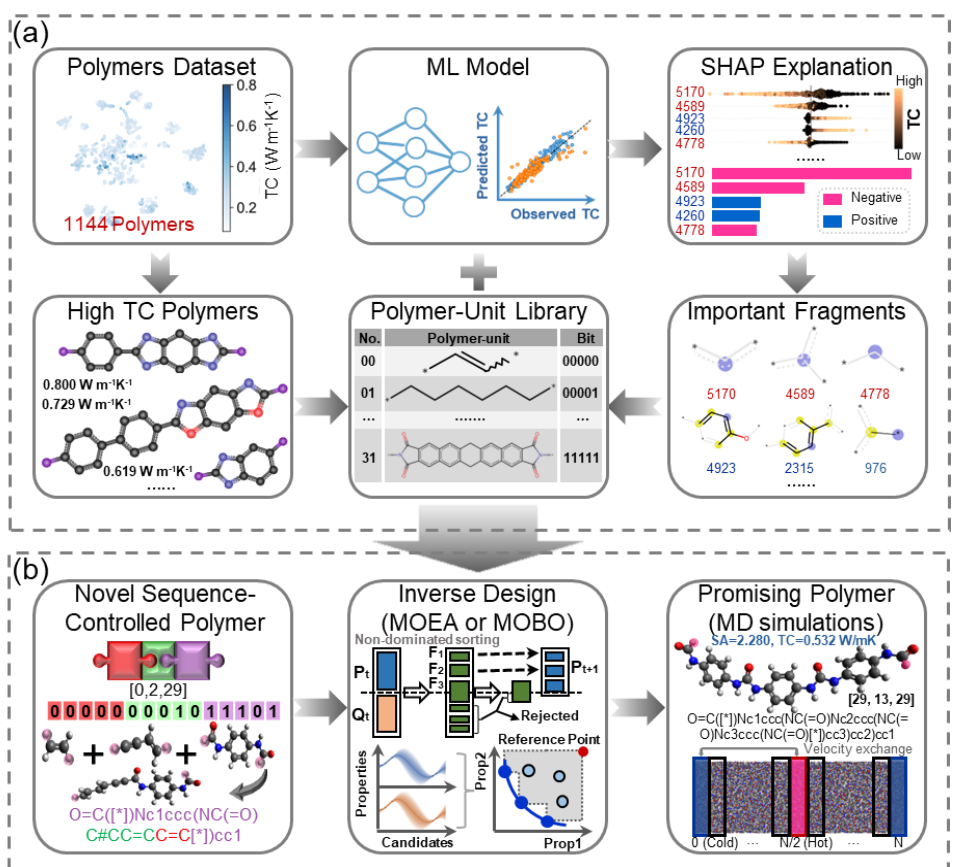

**Figure. 1.** Scheme for the design of sequence-controlled high thermal conductivity polymers. (a) ML model training and polymer-unit library generation. (b) Inverse design of polymers with high TC.

## 2 Methods

**2.1 Polymer representation and DNN surrogate model**

We trained a DNN surrogate model to predict the TC of polymers so as to maintain reasonable experimental costs. A polymer representation approach of MFF [43] was adopted to characterize the structure of polymers, which is an expansion of Morgan fingerprints to overcome the high dimensional limitations of vectors [44]. MFF has been successfully deployed in various tasks such as the discovery of multifunctional polyimides [15], the screening of innovative polymers for gas separation membranes [5], and the prediction of the free volume energy of polymer membranes [45]. In this work, MFF captures the frequency of chemical substructures with a radius of 3 units (each atom or bond is one unit) in 1144 polymer monomers. As a result, we counted 6926 chemical substructures from these 1144 polymers, of which the 194 most popular substructures with a frequency no less than 1100 times were retained as input features. More details of MFF can be found in the Supplementary Section A. For DNN model training, the 1144 polymer data were randomly split according to the training/testing set as 80%/20%, and the hyperparameters were optimized by KerasTuner [46] Toolkit with Adam optimizer, and mean squared error loss. The final DNN model has four hidden layers with 416, 256, 244 and 256 nodes, respectively; ReLU activation; and dropout of 0.5.

**2.2 Multi-objective optimization algorithms for polymer inverse design**

Two optimization algorithms of multi-objective optimization algorithms, unified non-dominated sorting genetic algorithm III (U-NSGA-III) [47] and q-noisy expected hypervolume improvement (qNEHVI) [48], which are MOEA and MOBO-based algorithms, respectively. U-NSGA-III is an updated version of NSGA-III [49,50], which improves the generalization of different dimensional objective problems by increasing the selection pressure through the introduction of a scalar selection operator. U-NSGA-III was implemented in the pymoo [51] package and kept all hyperparameters with default values. qNEHVI extends the acquisition function of expected improvement to hypervolume (HV) as an objective, and evaluates samples collected by the QMC sampler from the model posterior, which identifies the candidate with the largest objective value. The HV is the area enclosed by connecting the points at the Pareto front and a specified reference point in the bi-objective problem.[52] qNEHVI was operated in BoTorch [53] software and the base and raw sampling were set at 256 and 128, respectively, to speed up the computational runtime.

**2.3 Substructure contribution analysis using SHAP analysis**

The interpretable ML of the DNN model coupled with SHAP [41] analysis provides insights into the contribution of key input substructures to TC. SHAP is a game-theoretic approach that connects the optimal credit allocation of a model input features with local interpretations of the model [4]. The SHAP approach evaluates the performance of the ML model by ignoring each input feature sequentially and assigns a feature importance and the impact of each sample on the final prediction.

**2.4 Calculation of polymer properties**

Polymer modeling and MD simulations were performed in an automated computational framework, namely RadonPy [40], which is well integrated with several external chemical computation software such as RDKit [54] and LAMMPS [55]. RadonPy takes the SMILES of the polymer repeating units as input and reads in pre-defined parameters such as the polymerization degree of the individual chains and the number of polymer chains. In RadonPy, the generation of polymer single chains and the equilibration of amorphous systems are based on a self-avoiding random walk algorithm [56] and follow a 21-step equilibration scheme [57], respectively. Our study object is unified as an amorphous system containing 10 chains and ~10000 atoms. Once an equilibrium amorphous model was achieved, the *Rg* was calculated as follows:

$$R_g = \sqrt{\frac{1}{p}\sum_{i=1}^{p}(\boldsymbol{r}_i - \boldsymbol{r}_m)^2} \quad (1)$$

where $p$ is the degree of polymerisation of polymer chains, the $\boldsymbol{r}_i$ is the position of a repeating unit and $\boldsymbol{r}_m$ represents the mean position of the monomer in a polymer chain.

Afterwards, the equilibrium amorphous cell was replicated in triplicate along the *x*-direction (consistent with the direction of heat flux) under periodic boundary conditions, and the reverse NEMD simulation proposed by Müller-Plathe [58] was performed to calculate the TC. The NEMD simulation divides the simulation model into *N* blocks (*N*=20 in the *x*-direction) and periodically exchanges the velocity of the coldest atoms in the *N/2* block with that of the hottest atoms in the 0 block to create a temperature gradient. The TC of the polymer was solved using the following equation:

$$k = \frac{J}{\partial T/\partial x} = \frac{\Delta\delta}{2\Delta t(\partial T/\partial x)A} \quad (2)$$

where $J$ is the heat flux, $\partial T/\partial x$ is the temperature gradient, $\Delta t$ is the simulation time, $\Delta\delta$ is the exchanged energy and $A$ the cross-sectional area of the simulation box.

Ultimately, a decomposition analysis was carried out to quantify different contributions to TC, which are categorized into convective and non-convective effects according to the source of energy flux. Non-convective effects can be further dissected into the pairwise, bond, angle, dihedral, improper and K-space contributions. More details about polymer modelling and MD simulation are available in our previous work [59].

## 3 Results and Discussion

### 3.1 Polymer dataset, ML model and polymer-unit library

The training data containing 1144 polymers were collected from the PoLyInfo database, and their TCs were obtained by performing NEMD simulations of amorphous systems with ~ 30000 atoms using the RadonPy toolkit [40,59]. Considering the reasonable cost of this work, homopolymers were adopted as the research object, since RandnPy disclosed the MD-calculated TCs of more than a thousand homopolymers [40]. Moreover, the validation of the TC of some emerging homopolymers was performed using the same approach and parameters as calculated for these known TCs. The selected

6 / 22

polymers consist of over 20 types of backbones such as polyolefins, polyethers, polyimides and polyketones, which have been confirmed with good coverage of polymer structural features in the PoLyInfo database [40]. The distribution of TC is demonstrated in Fig. S2a, where most of the structures range from 0.1 - 0.4 W m$^{-1}$K$^{-1}$. Achieving intrinsically high TC is difficult, with only 4.63% of the polymers having a TC of > 0.4 W m$^{-1}$K$^{-1}$. These polymers were characterized in the form of the simplified molecular input line entry system (SMILES) [60] and transformed into MFFs for ML inputs [45]. MFFs have 194 dimensions, corresponding to the counts of the 194 most frequent substructures in the whole 1144 training dataset.

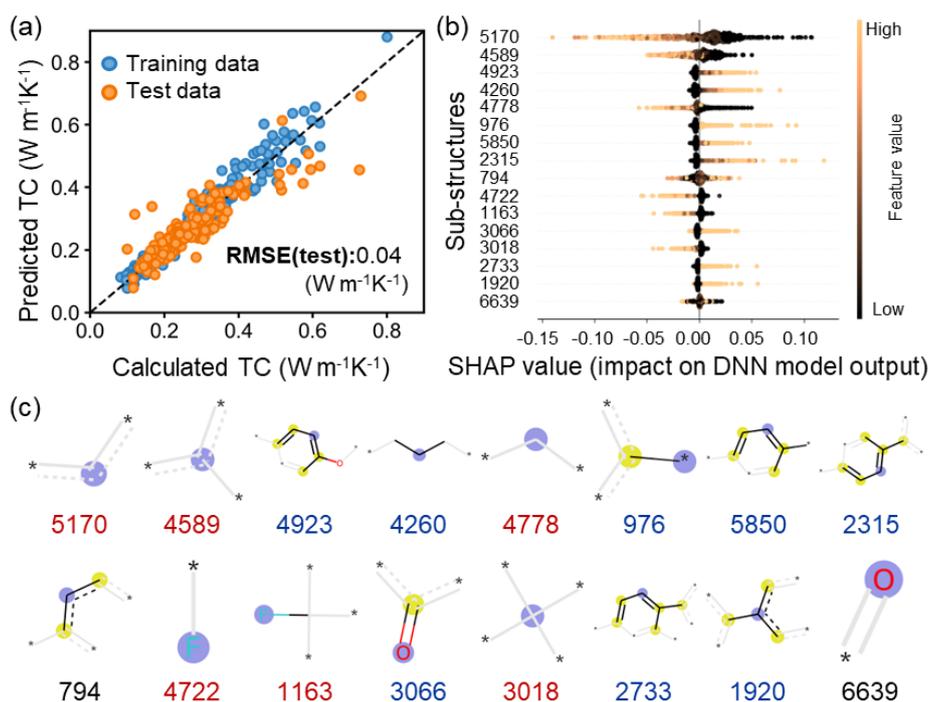

**Figure. 2.** ML model performance and feature importance evaluation. (a) ML result of DNN. (b) The interpretations of the DNN model for TC prediction by the SHAP evaluation. (c) The key sub-structures that act on TC, where blue text indicates a positive effect and red indicates an inhibitory effect.

We trained the DNN predictive model using the train/test ratio of 80%/20%, as shown in Fig. 2a. The ML-predicted TCs closely match those calculated by MD simulations, with a test root mean square error (RMSE) of 0.04 W m$^{-1}$K$^{-1}$. Apart from the DNN models, we additionally examined four ML models, namely random forest (RF), extreme gradient boosting (XGBoost), multi-layer perceptron (MLP), and Gaussian process regression (GPR), each of which was repeated 10 times with different training datasets. Fig. S2b-c summarize the test accuracies of the five ML models. Although these ML models have comparable capabilities, DNN is more stable and suitable as a surrogate model for TC simulation of new polymer structures. The SMILES of each emerging polymer was firstly transformed to MFF containing 194 bits and then fed into the trained DNN model to evaluate the TC.

**Table. 1.** Polymer fragments as basic units for high thermal conductivity polymer design. The structures of each polymer unit are displayed in Fig. S5, which were binary encoded according to serial numbers (No.).

| No. | SMILES of fragments | Code | No. | SMILES of fragments | Code |
|---|---|---|---|---|---|
| 0 | [*]C=C[*] | [00000] | 16 | [*]c1nc2cc3nc([*])[nH]c3cc2[nH]1 | [10000] |
| 1 | [*]CCCCCC[*] | [00001] | 17 | [*]CC(=O)N[*] | [10001] |
| 2 | [*]C#CC=C[*] | [00010] | 18 | [*]CNC(=O)N[*] | [10010] |
| 3 | [*]c1ccc([*])cc1 | [00011] | 19 | [*]C(=O)NNC([*])=O | [10011] |
| 4 | [*]c1ccc([*])[nH]1 | [00100] | 20 | [*]NNC(=O)C([*])=O | [10100] |
| 5 | [*]c1ccc2cc([*])ccc2c1 | [00101] | 21 | [*]c1ccc2oc([*])nc2c1 | [10101] |
| 6 | [*]c1ccc-2c(Cc3cc([*])ccc-23)c1 | [00110] | 22 | [*]c1nc2ccc([*])cc2o1 | [10110] |
| 7 | [*]CO[*] | [00111] | 23 | [*]NC(=O)C=CC(=O)N[*] | [10111] |
| 8 | [*]OC([*])=O | [01000] | 24 | [*]C(=O)C=CC(=O)N-[*] | [11000] |
| 9 | [*]c1ccc([*])o1 | [01001] | 25 | [*]NC(=O)c1ccc([*])cc1 | [11001] |
| 10 | [*]C(=O)C=CC([*])=O | [01010] | 26 | [*]Nc1ccc(C([*])=O)cc1 | [11010] |
| 11 | [*]C(=O)c1ccc(cc1)C([*])=O | [01011] | 27 | [*]N1C(=O)c2ccc([*])cc2C1=O | [11011] |
| 12 | [*]c1cnc([*])nc1 | [01100] | 28 | [*]NC(=O)c1ccc(cc1)C([*])=O | [11100] |
| 13 | [*]Nc1ccc(N[*])cc1 | [01101] | 29 | [*]C(=O)Nc1ccc(NC([*])=O)cc1 | [11101] |
| 14 | [*]c1nc2cc([*])ccc2[nH]1 | [01110] | 30 | [*]n1c(=O)c2cc3c(cc2c1=O)c(=O)n([*])c3=O | [11110] |
| 15 | [*]c1nc2ccc([*])cc2[nH]1 | [01111] | 31 | [*]N1C(=O)c2cc3cc4Cc5cc6cc7C(=O)N([*])C(=O)c7cc6cc5Cc4cc3cc2C1=O | [11111] |

To determine the polymer-unit library, we analyze the connection between substructures and TCs through SHAP. Fig. 2b illustrates the role of the most important 16 substructures on TC, where each dot indicates the effect of the substructure on the TC of an individual polymer. Based on the impact of different descriptor dimensions on the output of the DNN model, eight substructures play a positive role on TC in general, while six substructures inhibit it, which are marked with blue and red text in Fig. 2c, respectively. These structures coincide with insights extracted from our previous work on an ML model of polymer physical descriptors versus TC, i.e., that conjugated, linear side-chain-free polymers are favorable for maintaining large chain stiffness and thus maintaining high TC [59]. Moreover, when the polymer system contains heavy atoms such as F, it inhibits the effective transport of the heat flow thereby preventing the generation of high TC [10]. Combining our domain knowledge and the structural features of 53 high TC polymers (TC > 0.40 W m$^{-1}$K$^{-1}$, listed in Table S1 and Fig. S4), we constructed a polymer-unit library containing 32 small fragments listed in Table. 1. These base units consist of four atoms, C, H, O, as well as N, and were binary coded from [00000] to [11111] by sequential numbers to ensure the uniqueness of the identification for each fragment.

**3.2 Construction of triblock polymers database**

Ideally, once a polymer fragment library is identified, we could produce a dataset with an infinite number of polymers by adjusting the number and order of the polymer sequences, but we have to balance the synthesizability of the polymers, the cost of property simulation, and the actual hardware capabilities. We built a complete database of triblock polymers and calculated their TCs and SA scores for evaluating the performance of the MOEA and MOBO algorithms. The SA score was originally developed to characterize the synthesis accessibility of drug-like small molecules according to a combination of fragment contribution and complexity penalties, with values ranging from 1 (easy) to 10 (hard) [42]. Gradually, SA scores were migrated to the assessment of polymer synthesizability [38,61]. It is worth mentioning that Wu et al. [61] realized the synthesis of three easily processable polyimides by referring to SA scores. Fig. 3a provides a demo of triblock polymer formation, and each block is extracted from one of 32 possible sequences. Polymer sequences are directionless, for instance, a polymer consisting sequentially of units [0, 2, 29] is equivalent to one with [29, 2, 0], and their SMILES are O=C([*])Nc1ccc(NC(=O)C#CC=CC=C[*])cc1.

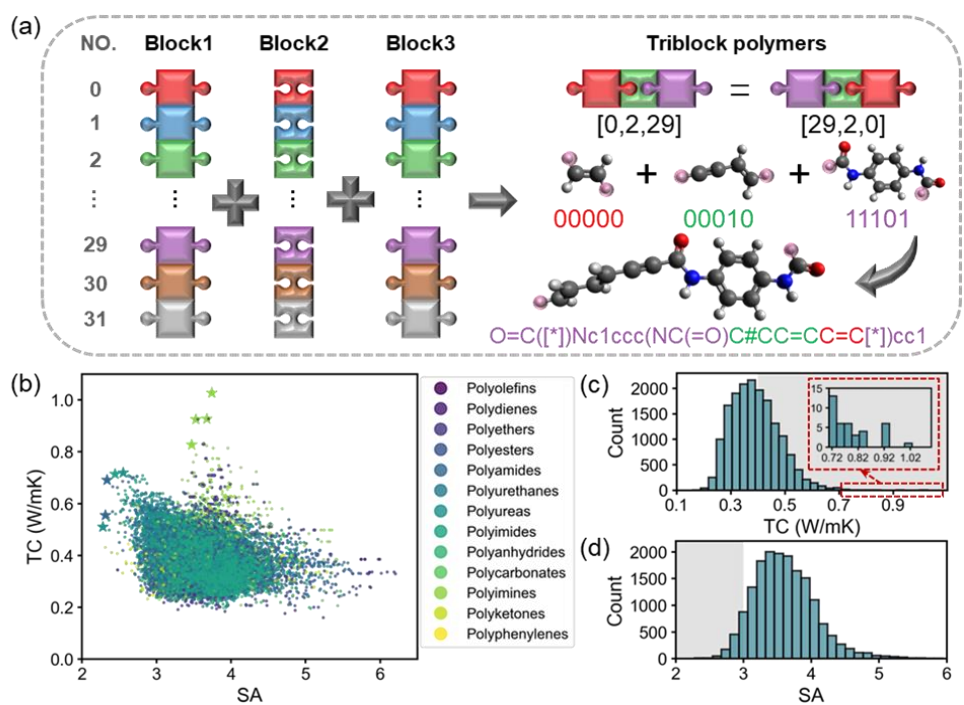

**Figure. 3**. Construction of triblock polymers dataset. (a) Example of the generation of a triblock polymer. (b) SA score versus TC of all 16896 triblock polymers, where stars indicate candidates at the Pareto front. (c) and (d) Distributions of the TC and SA for the whole triblock polymers. The gray backgrounds highlight the statistics of polymers with TC > 0.4 W m$^{-1}$K$^{-1}$ or SA < 3.0.

The relationships between TCs and SA scores of 16896 triblock polymers are illustrated in Fig. 3b. These polymers are classified into 13 categories referring to the same classification method as PoLyInfo, including polyolefins, polyethers and polyethers, etc. The DNN predicted TCs of candidates ranged from 0.16 to 1.03 W m$^{-1}$K$^{-1}$, of which 42.6% have TCs greater than 0.40 W m$^{-1}$K$^{-1}$ (see Fig. 3b). The SA scores in the range of 2.28 ~ 6.21, where 6.3% with SA scores less than 3.0. Nevertheless, it is even more difficult to achieve both high TC and low SA (TC > 0.4 W m$^{-1}$K$^{-1}$ and SA < 3.0) in a single polymer, with only 4.5% of candidates satisfying the requirements (ideal polymers). We recognized the Pareto front for the entire dataset, and there are nine candidates at the Pareto front, five of which are ideal polymers, while the rest only satisfy the characteristics of high TC (marked in Fig. 3b by stars).

**3.3 Performance evaluation for inverse design algorithms**

We compared two state-of-the-art multi-objective optimization algorithms of U-NSGA-III [47] and qNEHVI [48], and evaluated the optimization efficiency using the indicator of HV. Since U-NSGA-III in the pymoo software [51] was originally developed to investigate the minimization problems, we took a negative sign for the value of TC in each MOGA run, and the reference point was set as [0,-10] for TC and SA in turn. While the qNEHVI in the BoTorch package [53] was designed for maximization problems, we reversed the SA scores and used the reference point of [0,10]. Therefore, the largest HV is the area formed by the nine global Pareto optimal solutions with the reference point, which is 7.514.

Figures 4a-b exhibit the optimization trajectories for a single run of MOEA and MOBO with 10 random initial structures and 200 iterations × 10 candidates per batch, where nine gray stars mark the sites of global optimal polymers and the polymer dots are color-coded referring to the generations. The distribution of searched non-duplicated polymer structures in a MOBO run is much denser than those in a MOEA run. qNEHVI integrates HV into the expected improvement acquisition function as an objective to evaluate the randomized Quasi-Monte Carlo (QMC) samples sourced from the model posterior, and thus generates non-duplicated candidates in almost every generation [48]. This also enables the models to have the ability to break out of the local optimal solution and further makes the HV increase. The optimization strategy of U-NSGA-III is quite different, which is inspired by the behavior of genes in organisms that crossover and mutate during evolution, and the optimal polymers are designed by randomly selecting parents for matching and introducing a tournament operator [47]. However, the U-NSGA-III performance is affected by the initial polymer structures, as the optimization process is mainly an accumulation of previous polymer units with positive contributions, and therefore

it is easy to be trapped in the local optimal solution.

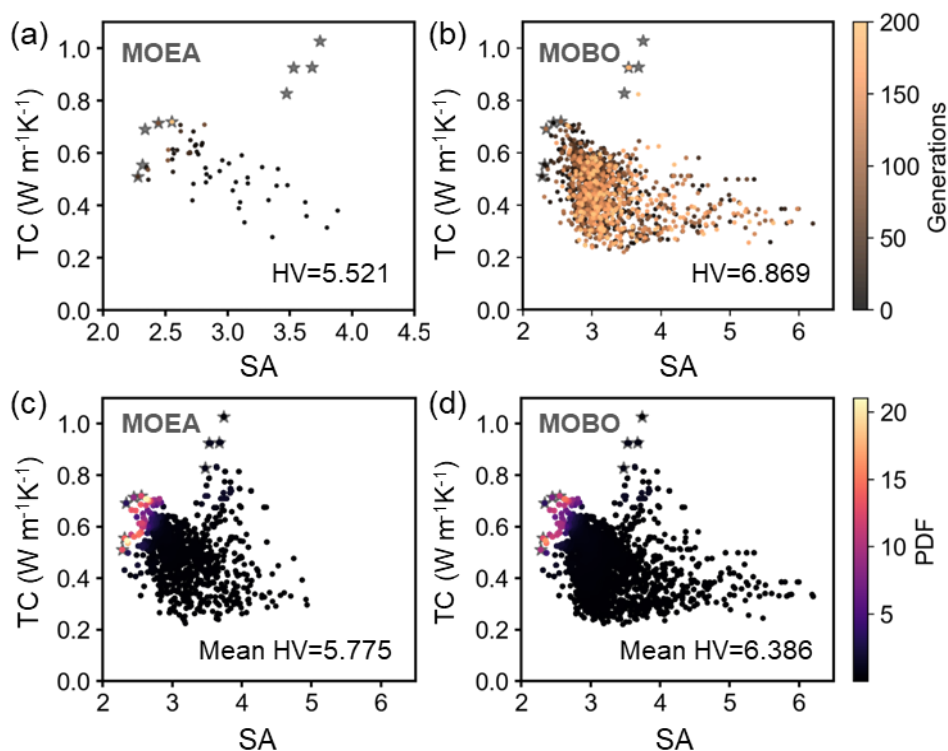

**Figure. 4.** Evaluation of multi-objective optimization algorithms. (a) and (b) Optimization trajectories for a single run of MOEA and MOBA with 10 random initial structures and 200 iterations × 10 candidates per batch. (c) and (d) Probability density maps in objective space for 20 runs of MOEA and MOBA, respectively.

For obtaining statistical results we performed 20 runs of the MOEA and MOBO algorithms with different initial candidates, respectively, and the HV convergence curves are displayed in Fig. S6a-b. HVs of U-NSGA-III can rapidly rise to a certain level (within 20 generations), but it is difficult to increase again in subsequent. However, there are three qNEHVI runs that identified nine global optimal polymers within 200 generations and almost all of the HVs get a secondary boost after the first time to a certain level. The difference in this enhancement depends on the stochastic nature of QMC sampling [62]. All the HVs of optimization algorithms reach a referred value that is calculated by the five ideal global optimal Pareto polymers and the referred point, although the mean HV of MOBO is greater than that of MOEA (see Fig. S6c-d). Our work aims to explore as many promising polymers as possible (TC > 0.4 W m$^{-1}$K$^{-1}$ and SA < 3.0), we employed the Gaussian kernel to estimate the probability density function (PDF) of all searched polymers in 20 MOEA and MOBO with various random starts, as shown in Fig. 4d-c, separately. The high probability region in both maps occurs close to the five ideal polymers at the global Pareto front, which reflects the robustness of the two optimization algorithms. Compared to U-NSGA-III, there are more qNEHVI-searched polymers far from the global Pareto front, due to uncertainties in the QMC sampler. Overall, MOBO can keep the diversity of optimized polymers along the Pareto front, whilst this also requires more QMC sampling attempts and higher experimental costs. MOEA is capable of efficiently and economically converging to an optimized solution, but the gap of TC enhancement is affected by the initial candidates. The additional discussions of the influence of the initial structures on the convergence performance of MOEAs are given in Supplementary Section E.

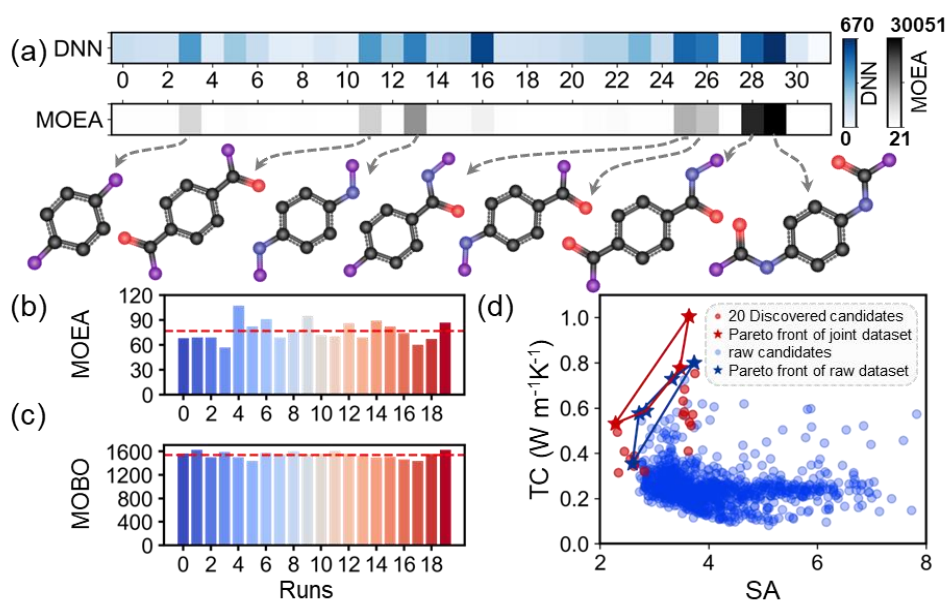

**Figure. 5.** Statistics of high-frequency polymer units in 20 MOEA runs and generation of promising triblock polymers. (a) Genetic strips show the frequency of occurrence of polymer units, where the grayish-white strip was based on an ensemble of 20 MOEA optimization runs, and the blue-white strip from the 2542 polymers with DNN-predicted TC≥0.50 W/mK or calculated SA≤3.0. (b) and (c) Number of candidates designed by MOEA and MOBO after de-duplication in 20 runs. (d) Pareto front improvement over the 1144 raw training data after adding 20 MOEA-optimized candidates with MD-calculated TC.

We extracted the frequency of occurrence of polymer units in 20 runs of MOEA optimization to capture the contribution of different fragments to TC, as depicted by the grey-white strip in Fig. 5a. The top seven fragments with the largest frequencies are all aromatic fragments containing benzene rings, where the top-ranked polymer unit is [*]C(=O)Nc1ccc(NC([*])=O)cc1, with 30051 occurrences. The MOEA-recommended polymer units are in close agreement with the statistics from the ideal polymers with DNN-predicted TCs (blue-white stripe derived from the statistics of fragments in 2542 polymers with TC $\geq$ 0.50 W m$^{-1}$K$^{-1}$ or SA $\leq$ 3.0). It reflects that the MOEA algorithm has excellent optimization performance, and assists in the rapid identification of promising polymer units. Figure. 5b and c outline the number of explored polymers (de-duplicated) in 20 MOEA and MOBO runs, respectively. The effective number of polymers per MOEA cycle is much less than that of MOBO, with a mean value of about 77, which is less than 5.0% of the average value for MOBO. Therefore, an effective scheme is the design of high TC polymers through multiple parallel MOEAs with different random states, so as to reduce the impact of the initial structures. In addition, we calculated the thermal conductivity of 20 MOEA-designed polymers (red dots) using NEMD in Fig. 5d, which indeed improves the Pareto front (marked by stars) formed with 1,144 raw polymers (blue dots).

### 3.4 Inverse design of pentablock polymers

We operated 20 parallel MOEA algorithms to design high TC pentablock polymers in a vast space of more than ten million candidates. Figure 6a statistics the HV raising curves for 20 MOEAs, where each MOEA run started with 10 random structures and went through 200 iterations × 10 candidates per batch. After 200 generations, the HVs of 20 MOEAs range from 6.30 to 6.95. The parallel scheme compensates to some extent for the fact that the performance of the genetic algorithm is limited by the initial structures, thus exploring more polymers that satisfy the target properties. Moreover, the number of effective polymers developed in all 20 runs is below 130 (Insert in Fig. 6a), with a total of 1921 non-repeating polymers in the end. The value is smaller than the number of polymers produced by a MOBO (2005 non-repeating polymers) with a random state at the same conditions, revealing that parallel MOEAs are still capable of maintaining a reasonable experimental cost.

The pair plot of SA and TC for 1921 MOEA-derived polymers is exhibited in Fig. 6b, where more than half of the candidates satisfy predefined requirements, i.e., SA≤3.0 and TC≥0.40 W $m^{-1}K^{-1}$. However, only 338 of 2005 polymers meet the above conditions in a MOBO run, as displayed in Fig. S11. Considering TC and SA individually in Fig. 5c-f, the majority of polymers (above 86.4%) have a TC greater than 0.40 W $m^{-1}K^{-1}$ in parallel MOEA runs, whereas the proportion is only 46.1% in a MOBO runs. Similarly, MOEA runs have more polymers with SA scores of no more than 3.0 compared to the outcomes of a MOBO run, accounting for 57.5% and 18.6%, respectively. Parallel MOEAs scheme compensates for the lack of genetic algorithms limited by the initial structures, and is superior to MOBO at a comparable experimental cost. In addition, it is worth emphasizing that we use a DNN model to simulate the TC of polymers, and the prediction error of the model may lead to bias in the direction of optimization. Of course, this can be settled by using techniques such as MD simulations or experiments instead of ML surrogate models to calculate the properties of polymers in realistic applications.

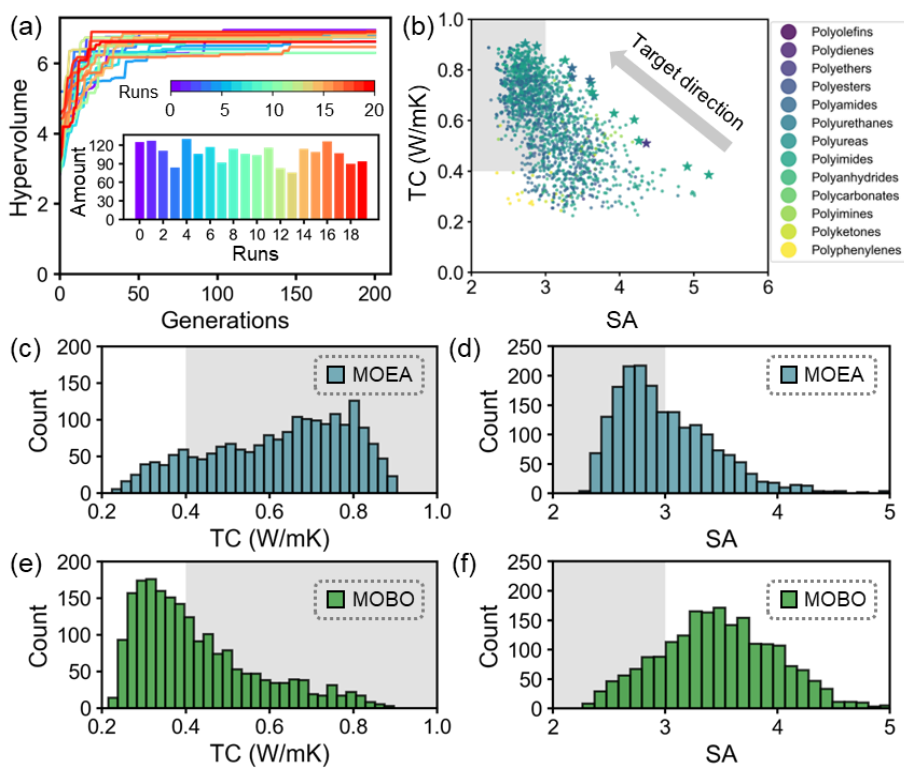

**Figure. 6.** Design of pentablock polymers with high TC. (a) Learning curves for 20 MOEA optimization runs with 10 different initial structures, and 200 iterations × 10 candidates per batch. (b) Ensemble of polymers generated by 20 MOEA optimization runs. (c) and (d) Distribution of TC and SA of 1921 non-repeating polymers obtained by 20 MOEA runs. (e) and (f) Distribution of TC and SA of 2005 non-repeating polymers obtained by a MOBO run with 10 different initial structures, and 200 iterations × 10 candidates per batch.

### 3.5 Insights into the linkage between polymer chain conformation and TC

The TC of polymers is closely linked to their microstructures, and the radius of gyration ($R_g$) was adopted to characterize the chain morphology in amorphous systems. We selected 50 MOEA-designed polymers, 20 of which are triblock polymers and 30 of which are pentablock polymers, and calculated their $R_g$ and TC through MD simulations, as shown in Fig. 7a (more details about 50 polymers are listed in Table S2). The $R_g$ of polymers exhibits a positive correlation with TC, since a large $R_g$ indicates that the polymer has strong intra-chain interactions, which facilitates heat transport across the amorphous system [59]. Furthermore, the decomposition analysis was implemented to understand the thermal transport mechanism, whereby the contributions of the TC were quantified into six components relating to convection, bond, angle, dihedral, improper and nonbonded. The nonbonded term was described as pairwise and K-space contributions. Figure 7b outlines six high TC polymers using decomposition analysis, of which half are triblock polymers (Fig. 7c) and half are pentablock polymers (Fig. 7d). All six candidates are conjugated aromatic polymers, and the structure with the highest TC is pentablock polymer (Pen_01) consisting of four benzene rings and one naphthalene ring. The benzene ring and its derived aromatic rings exhibit favourable structural stability and rigidity due to features such as coplanarity of the atoms and sp2 hybridization of the carbon atoms. It is clear that rigid monomers are a prerequisite for realizing high TC in amorphous systems, which is accompanied by a dominant contribution to TC from intrachain heat transport caused by bond, angle and dihedral interatomic interactions. In addition, we compared the structures of tri- and pentablock polymers, and the triblock polymers are more easily able to achieve large thermal conductivities, whilst the pentablock polymers have a higher possibility of synthesis owing to the longer monomer sequences. Multi-objective optimization algorithms are capable of designing polymers with excellent performance by comprehensively evaluating multiple factors of properties and synthesis.

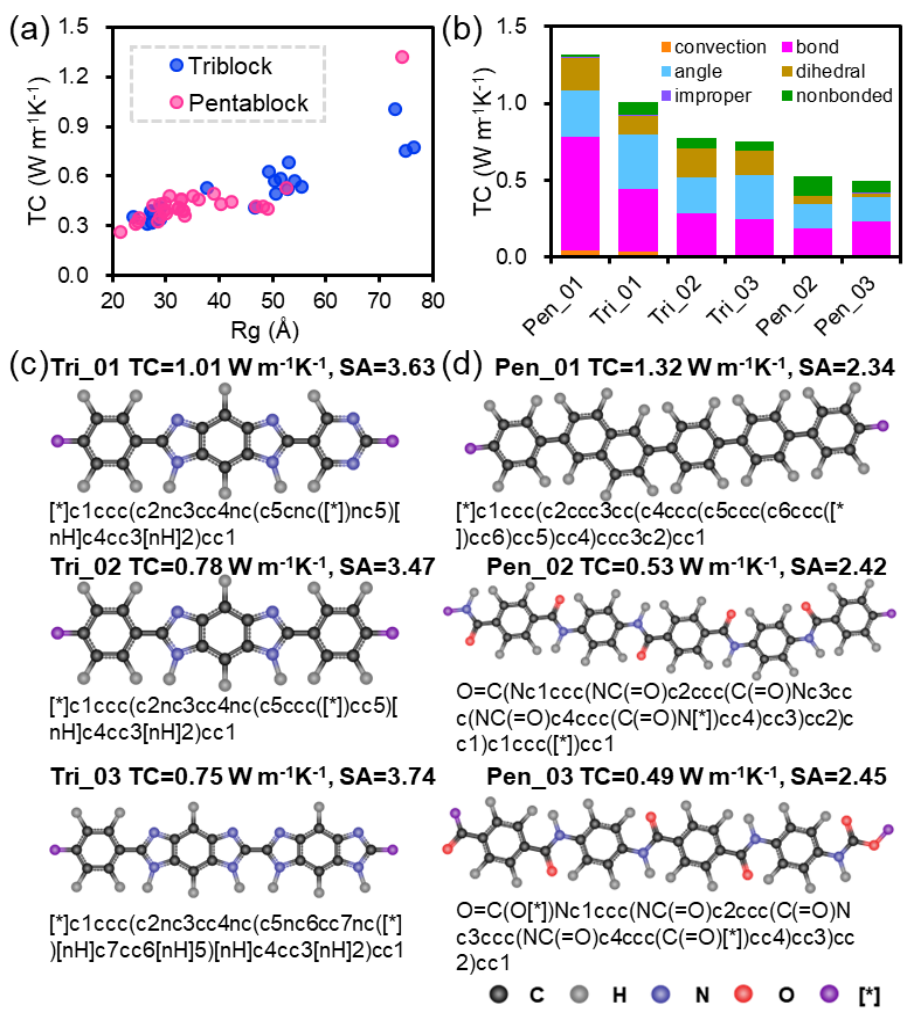

**Figure. 7.** Linkages between polymer chain conformation and TC. (a) Radius of gyration versus TC. (b) Quantitative decomposition of TC into contributions from convection and different types of interactions of six high TC polymers, where half are triblock (Tri) polymers, and others are pentablock (Pen) polymers, as shown in (c) and (d).

# 4 Conclusions

In conclusion, we have developed a robust AI-assisted framework for the inverse design of high intrinsic thermal conductivity polymers. We started with a computational dataset of 1144 polymers with MD-calculated TCs and constructed a DNN model to establish the relationships between monomer structures and TCs. The DNN model was not only utilized for TC evaluation of emerging designed polymers, but also guided the role of key chemical features on TC together with the SHAP analysis. Referring to the SHAP outputs and our domain knowledge, we built a polymer-unit library with 32 fragments and encoded them binary as [00000] to [11111].

We set our goal of designing target polymers with TC≥0.40 W m$^{-1}$K$^{-1}$ and SA≤3.0, since the synthesis possibilities of the polymers were also evaluated simultaneously. We then compared the two optimization algorithms of U-NSGA-III (MOEA) and qNEHVI (MOBO) in the entire dataset of triblock polymers produced by recombination with 32 polymer-unit sequences. Our results suggest that qNHEVI is capable of exploring a diversity of optimal polymers at the Pareto front, but the uncertainty in QMC sampling makes the trials costly. The performance of U-NSGA-III is affected by the initial random structures and usually falls into a locally optimal solution, but it has a clear low-cost advantage. Therefore, we performed 20 parallel MOEAs with various random states for the design of high thermal conductivity pentablock polymers. Among the 1921 generated polymers, more than half satisfy the predefined goal, superior to the results from a MOBO run. Finally, we calculated the TC of 50 newly designed polymers using MD simulations and probed a closely positive correlation between the $R_g$ of the chains and the TC in the amorphous systems. Further, by analyzing six polymers with high TC, all of which have a benzene ring-containing conjugated structure with large chain stiffness and strong intra-chain thermal transport.

The proposed ML-assisted design framework is universal and allows for generalization to other property targets including refractive index (RI), band gap, dielectric constant, glass transition temperature, and so on. First, the polymer-unit library is user-friendly and supports customization. Chemical blocks can be identified with domain knowledge and specific optimization targets, as well as better balancing additional constraints such as synthesizability, toxicity, and cost. For example, high thermal conductivity polymers are not favorable to heavy atoms, so it is possible to limit the chemical elements with atoms such as carbon, hydrogen, oxygen and nitrogen. Secondly, the developed parallel MOEAs do not require tedious hyperparameters tuning compared to generative algorithms such as VAE and RNN, and have the advantages of being lightweight, efficient and low-cost. Moreover, the fitness functions of the algorithms are variable to match different optimization objectives. An extended case on the design of innovative triblock polymers with TC > 0.40 W m$^{-1}$K$^{-1}$ and RI > 1.80 in Supplementary Section H confirms this point.

Going forward, we expect to generalize this scheme for more complex polymer systems. On the one hand, efforts are made to increase the accuracy of property evaluation methods (ML or experimental, etc.). On the other hand, a suitable polymer synthesis scoring function can be further established by integrating chemical reaction rules [63] or natural language processing [27,38].

## Credit author statement

All the authors have given approval to the final version of the manuscript. S. J. Initiated and concepted this research project; X. H. prepared the data and code, conducted the research, performed all simulations and wrote this manuscript under the guidance of S. J., C.Z. and H.W. supervised the project.

## Declaration of competing interest

The authors declare no conflicts of interest.

## Acknowledgements

This work was supported by Shanghai Key Fundamental Research Grant (No. 21JC1403300), Shanghai Pujiang Program (No. 20PJ1407500), and the National Natural Science Foundation of China (No. 52006134),. The computations in this paper were run on the π 2.0 cluster supported by the Center for High Performance Computing at Shanghai Jiao Tong University.

## Data Availability

The high-fidelity DNN surrogate model and the codes for the inverse design of high thermal conductivity polymers are available in the GitHub repository: https://github.com/SJTU-MI/Inverse_Design_of_Polymers. Detailed descriptions can be found in the Methods Section and Supplementary material.

# Supporting Information for

AI-assisted inverse design of sequence-ordered high intrinsic thermal conductivity polymers

Xiang Huang[a], C. Y. Zhao[a], Hong Wang[b], and Shenghong Ju[a, b, *]
[a] China-UK Low Carbon College, Shanghai Jiao Tong University, Shanghai, China
[b] Materials Genome Initiative Center, School of Material Science and Engineering, Shanghai Jiao Tong University, Shanghai, China

* Corresponding email: shenghong.ju@sjtu.edu.cn.

**This PDF file includes:**

    Supporting text
    Figures S1 to S14
    Tables S1 and S2
    SI References

**Supporting Information Text**

**A. Morgan fingerprints with frequency (MFF) for polymer representation**

Morgan fingerprints with frequency (MFF) is an expansion of Morgan fingerprints to overcome the high dimensional limitations of vectors [1]. In the Morgan algorithm, different atoms (depending on the type and nearest neighbors under a predefined radius) are given unique hashed identifiers (bit vectors). Thus, each identifier corresponds to a specified substructure. For the benchmark dataset, we can count the number and frequency of different identifiers and set a frequency threshold to determine the composition of the MFF vector. In this work, we counted 6926 chemical substructures with radius of 3 from the repeating units of 1144 polymers, of which the 194 most popular substructures with a frequency no less than 1100 times were retained as input features. For example, a monomer with PI813 (Figure S1a) has 32 types of substructures, where parts of which are shown in Figure S1b. Moreover, 15 substructures are valid, corresponding to the 194 most frequent segments. In Figure S1c, the positions of the 15 substructures were sequentially assigned frequencies (integers), i.e., the MFF for the PI813.

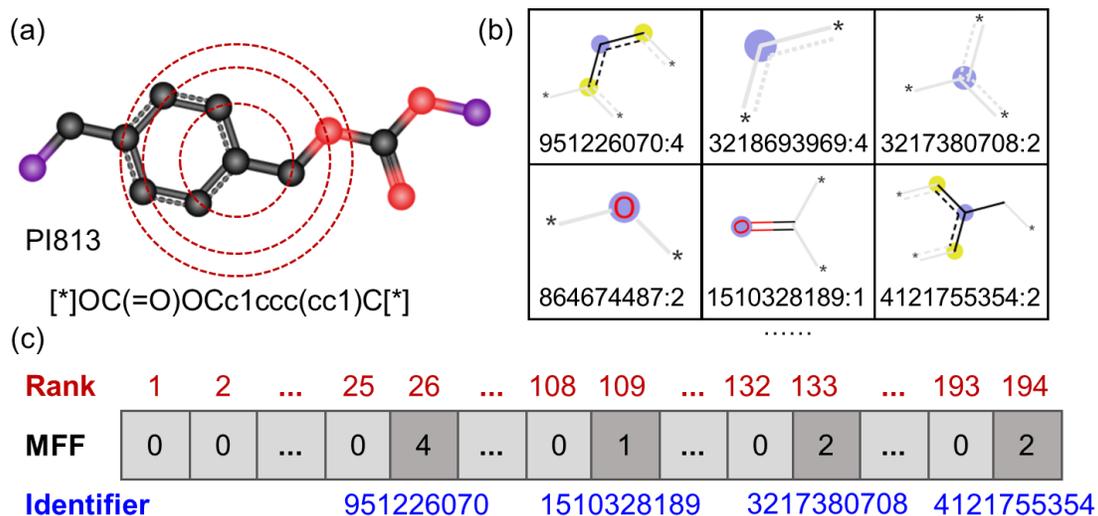

**Fig. S1.** Example of MFF generation. (a) Repeating unit of polymer with ID of PI813. (b) Library of polymer substructures. (c) MFF vector

## B. Machine learning models trained by 1144 polymers with known thermal conductivity

We evaluated the performance of five machine learning (ML) algorithms of deep neural networks (DNN), random forests (RF), eXtreme gradient boosting (XGBoost), multilayer perceptron (MLP), and Gaussian process regression (GPR) separately. The RF, XGBoost and MLP were executed in the Scikit-learn repository with 10-fold cross-validation [2], and the GPR was repeated 10 runs using different training sets in the Gpytorch toolkit [3]. Figure S2a statistics of polymer thermal conductivity (TC) distribution based on Gaussian kernel density estimation. The thermal conductivity of polymers is mainly distributed in the range of 0.1-0.4 W m$^{-1}$K$^{-1}$, and achieving a high thermal conductivity is quite difficult, with only 4.63%. The root-mean-square error (RMSE) and R-square ($R^2$) for five ML models are shown in Figure S2b and c. Overall, the performance of the five models is comparable, but the DNN model is more stable in 10 repetitions of the trial.

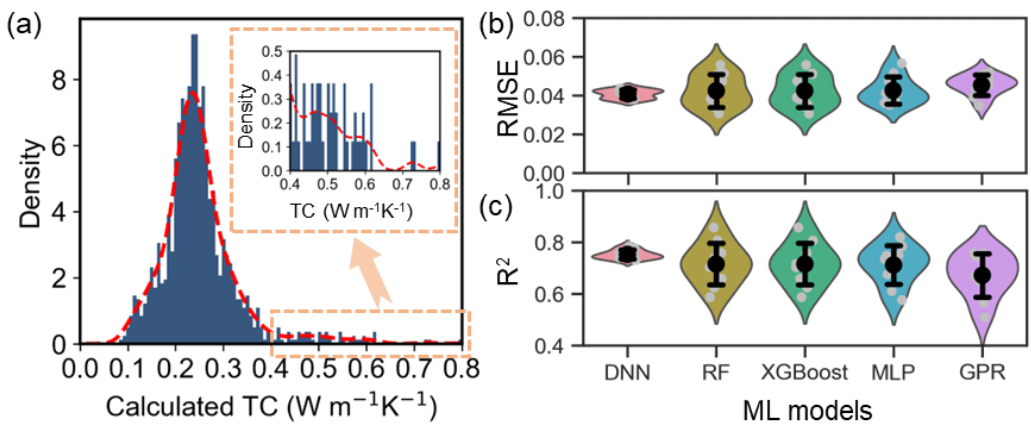

**Fig. S2.** Evaluation of machine learning models trained by 1144 polymer data. (a) Kernel density estimate plot visualizes the distribution of TC among 1144 polymers. (b) and (c) Root-mean-square error (RMSE) and R-square ($R^2$) for deep neural networks (DNN), random forests (RF), eXtreme gradient boosting (XGBoost), multilayer perceptron (MLP), as well as Gaussian process regression (GPR) models. To acquire statistical results, each model was repeated 10 runs using different training sets. Violins represent the distributions of the subsampling results, mean and standard deviation of MSE are shown in black, and individual subsample results are in gray.

The training and test $R^2$ of the surrogate DNN model are 0.95 and 0.79, respectively. We performed an additional five-fold cross-validation (CV) to evaluate the accuracy of the DNN model, as shown in Figure S3. In the five-fold cross-validation, the training $R^2$ of the DNN models ranged from 0.90 to 0.97, and the test $R^2$ ranged from 0.67 to 0.76. The accuracy of the surrogate prediction model utilized in this work is basically consistent with the results from five-fold cross-validation, reflecting the fact that the prediction model is reliable.

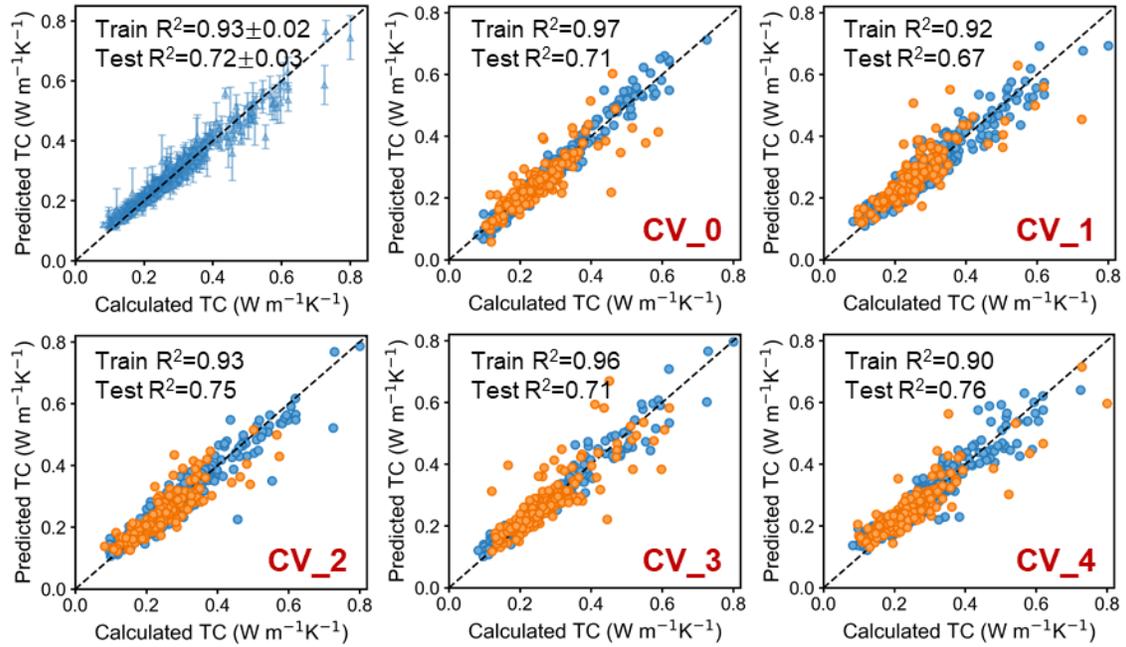

**Fig. S3.** Performance evaluation of DNN models based on five-fold cross-validation

## C. Demonstration of known high thermal conductive polymers and construction of polymer-unit library

Among the 1144 training data, 53 polymers have thermal conductivities exceeding 0.40 W m$^{-1}$K$^{-1}$, as listed in Table S1 and shown in Figure S4. Combining the structural characteristics of highly thermal conductive polymers and the outcomes from the shapley additive explanations (SHAP) [4], a polymer-unit library was generated as shown in Figure S5.

**Table S1.** List of amorphous polymers with high intrinsic thermal conductivity (TC > 0.40 W m$^{-1}$K$^{-1}$), where the corresponding monomer structures are illustrated in Figure S4.

| ID | SMILES | TC (W m$^{-1}$K$^{-1}$) |
| --- | --- | --- |
| AP1 | [*]c1ccc(-c2nc3cc4nc([*])[nH]c4cc3[nH]2)cc1 | 0.800 |
| AP2 | [*]c1ccc(-c2ccc(-c3nc4cc5nc([*])oc5cc4o3)cc2)cc1 | 0.729 |
| AP3 | [*]c1ccc(-n2c(=O)c3cc4c(=O)n([*])c(=O)c4cc3c2=O)cc1 | 0.725 |
| AP4 | [*]c1ccc(-n2c(=O)c3cc4c(=O)n([*])c(=O)c4cc3c2=O)c(C)c1 | 0.619 |
| AP5 | [*]c1ccc2[nH]c([*])nc2c1 | 0.619 |
| AP6 | [*]c1ccc(-c2nc3cc4nc([*])oc4cc3o2)c(O)c1 | 0.618 |
| AP7 | [*]NC(=O)C=CC(=O)Nc1nc([*])nc(N)n1 | 0.606 |
| AP8 | [*]c1ccc([*])s1 | 0.597 |
| AP9 | [*]C1=CC2=NC([*])=CC2=N1 | 0.594 |
| AP10 | [*]Nc1ccc(C#Cc2ccc(NC(=O)c3ccc(C([*])=O)cc3)cc2)cc1 | 0.588 |
| AP11 | [*]c1ccc(-c2ccc(-n3c(=O)c4cc5c(=O)n([*])c(=O)c5cc4c3=O)c(C)c2)cc1C | 0.581 |
| AP12 | [*]NC(=O)c1ccc(cc1)C(=O)Nc1ccc(cc1)c1ccc(cc1)[*] | 0.576 |
| AP13 | [*]/C=C\[*] | 0.573 |
| AP14 | [*]c1ccc2c(c1)Cc1cc(-n3c(=O)c4cc5c(=O)n([*])c(=O)c5cc4c3=O)ccc1-2 | 0.566 |
| AP15 | [*]c1ccc2cc(-c3nc4ccc(-c5ccc6nc([*])[nH]c6c5)cc4[nH]3)ccc2c1 | 0.553 |
| AP16 | [*]c1nc2cc3nc(-c4ccc([*])o4)[nH]c3cc2[nH]1 | 0.547 |
| AP17 | [*]Nc1ccc(NC(=O)c2ccc(C([*])=O)cc2)cc1 | 0.545 |
| AP18 | [*]c1ccc(-c2ccc(-c3nc4ccc(-c5ccc6nc([*])oc6c5)cc4o3)cc2)cc1 | 0.542 |
| AP19 | [*]Nc1ccc(C([*])=O)cc1 | 0.527 |
| AP20 | [*]C=C([*])F | 0.522 |
| AP21 | [*]c1ccc([*])[nH]1 | 0.517 |
| AP22 | [*]c1ccc(-c2ccc(-n3c(=O)c4cc5c(=O)n([*])c(=O)c5cc4c3=O)c(OC)c2)cc1OC | 0.515 |
| AP23 | [*]c1c(C)c(C)c(-n2c(=O)c3cc4c(=O)n([*])c(=O)c4cc3c2=O)c(C)c1C | 0.514 |
| AP24 | [*]c1ccc(-c2ccc(-c3ccc(N4C(=O)c5ccc(-c6ccc7c(c6)C(=O)N([*])C7=O)cc5C4=O)cc3)cc2)cc1 | 0.511 |
| AP25 | [*]N1C(=O)c2c(C1=O)cc(cc2)c1cc2c(C(=O)N(C2=O)c2ccc(cc2)c2ccc(cc2)[*])cc1 | 0.509 |
| AP26 | [*]NNC(=O)C([*])=O | 0.504 |
| AP27 | [*]C(O)C([*])O | 0.503 |
| AP28 | [*]c1ccc(C(=O)Nc2ccc(-n3c(=O)c4cc5c(=O)n([*])c(=O)c5cc4c3=O)cc2)cc1 | 0.501 |
| AP29 | [*]NC(=O)c1ccc(C(=O)Nc2cnc([*])nc2)cc1 | 0.491 |
| AP30 | [*]c1ccc(N2C(=O)c3ccc(-c4ccc5c(c4)C(=O)N([*])C5=O)cc3C2=O)nc1 | 0.483 |
| AP31 | [*]c1ccc(N2C(=O)c3ccc(-c4ccc5c(c4)C(=O)N([*])C5=O)cc3C2=O)cc1 | 0.482 |
| AP32 | [*]Nc1ccc(NC(=O)C=CC([*])=O)cc1 | 0.479 |
| AP33 | [*]NC(=O)c1ccc(cc1)C(=O)Nc1ccc(cc1)[*] | 0.473 |
| AP34 | [*]c1ccc(-c2nc3cc(-n4c(=O)c5cc6c(=O)n([*])c(=O)c6cc5c4=O)ccc3[nH]2)cc1 | 0.472 |

| ID | SMILES | TC (W m$^{-1}$K$^{-1}$) |
|---|---|---|
| AP35 | [*]CNC(=O)N[*] | 0.470 |
| AP36 | [*]c1ccc(NC(=O)c2ccc(C(=O)Nc3ccc(-n4c(=O)c5cc6c(=O)n([*])c(=O)c6cc5c4=O)cc3)cc2)cc1 | 0.468 |
| AP37 | [*]c1ccc(-c2ccc(N3C(=O)c4ccc(-c5ccc6c(c5)C(=O)N([*])C6=O)cc4C3=O)c(C)c2)cc1C | 0.463 |
| AP38 | [*]c1ccc2c(c1)Cc1cc(N3C(=O)c4ccc(-c5ccc6c(c5)C(=O)N([*])C6=O)cc4C3=O)ccc1-2 | 0.461 |
| AP39 | [*]NNC(=O)C=CC(=O)Nc1ccc(NC(=O)C=CC([*])=O)cc1 | 0.460 |
| AP40 | [*]CC[*] | 0.456 |
| AP41 | [*]Nc1nnc(Nc2n[nH]c([*])n2)[nH]1 | 0.451 |
| AP42 | [*]CC([*])C(N)=O | 0.445 |
| AP43 | [*]C(=O)Nc1ccc(NC(=O)c2ccc(N3C(=O)c4ccc([*])cc4C3=O)cc2)cc1 | 0.440 |
| AP44 | [*]C(C[*])O | 0.439 |
| AP45 | [*]c1ccc2oc([*])nc2c1 | 0.436 |
| AP46 | [*]c1ccc(Oc2ccc(N3C(=O)c4cc5C(=O)N([*])C(=O)c5cc4C3=O)cc2)cc1 | 0.425 |
| AP47 | [*]c1ccc(Oc2ccc(-c3ccc(Oc4ccc(-n5c(=O)c6cc7c(=O)n([*])c(=O)c7cc6c5=O)cc4)cc3)cc2)cc1 | 0.419 |
| AP48 | [*]NNC(=O)c1ccc(-c2ccc(C(=O)NNC(=O)c3cccc(C([*])=O)c3)cc2)cc1 | 0.419 |
| AP49 | [*]c1ccc(Nc2ccc(-n3c(=O)c4cc5c(=O)n([*])c(=O)c5cc4c3=O)cc2)cc1 | 0.419 |
| AP50 | [*]C(=O)NNC(=O)c1ccc([*])nc1 | 0.415 |
| AP51 | [*]c1ccc(-c2nc3cc(-c4ccc5oc([*])nc5c4)ccc3o2)cc1 | 0.410 |
| AP52 | [*]c1ccc(C(=O)Oc2ccc(-n3c(=O)c4cc5c(=O)n([*])c(=O)c5cc4c3=O)cc2)cc1 | 0.403 |
| AP53 | [*]c1ccc(-c2ccc(N3C(=O)c4ccc(-c5ccc6c(c5)C(=O)N([*])C6=O)cc4C3=O)c(OC)c2)cc1OC | 0.401 |

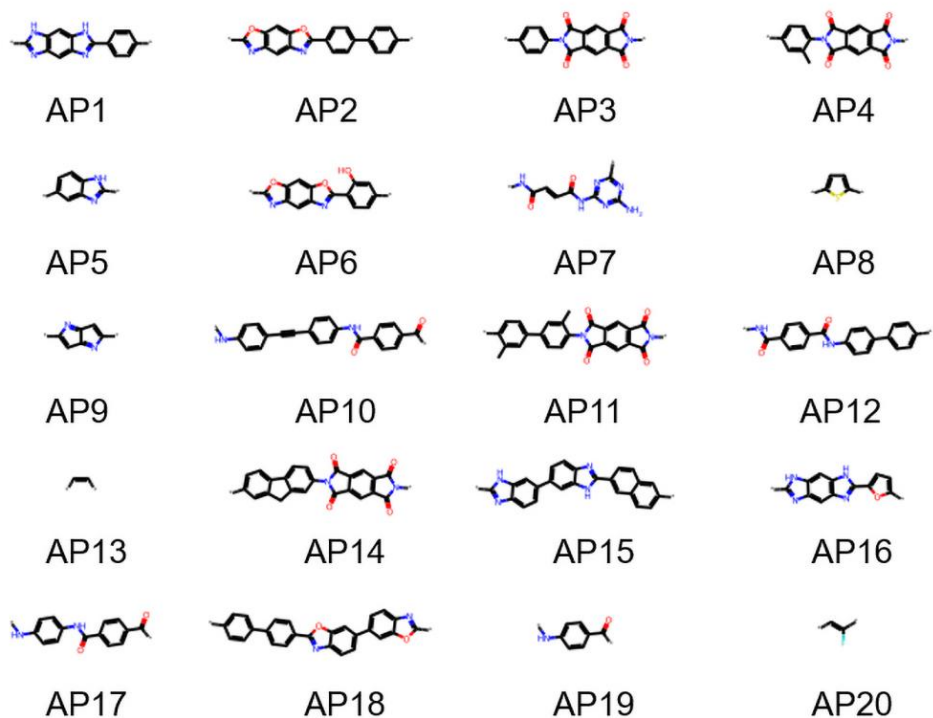

**Fig. S4.** Continued.

S6

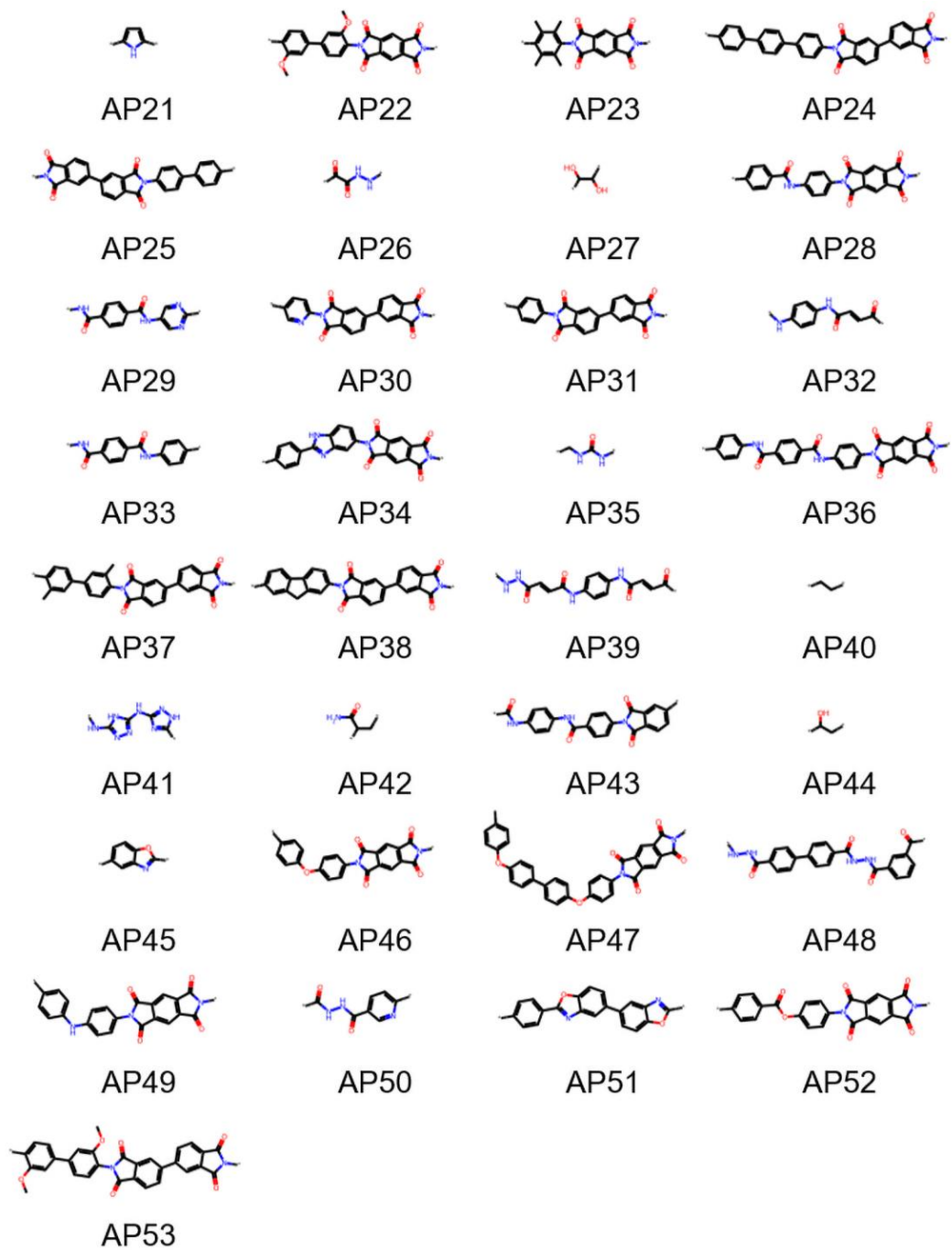

**Fig. S4.** Structures of polymer repeating units with thermal conductivity greater than 0.40 W/(mK).

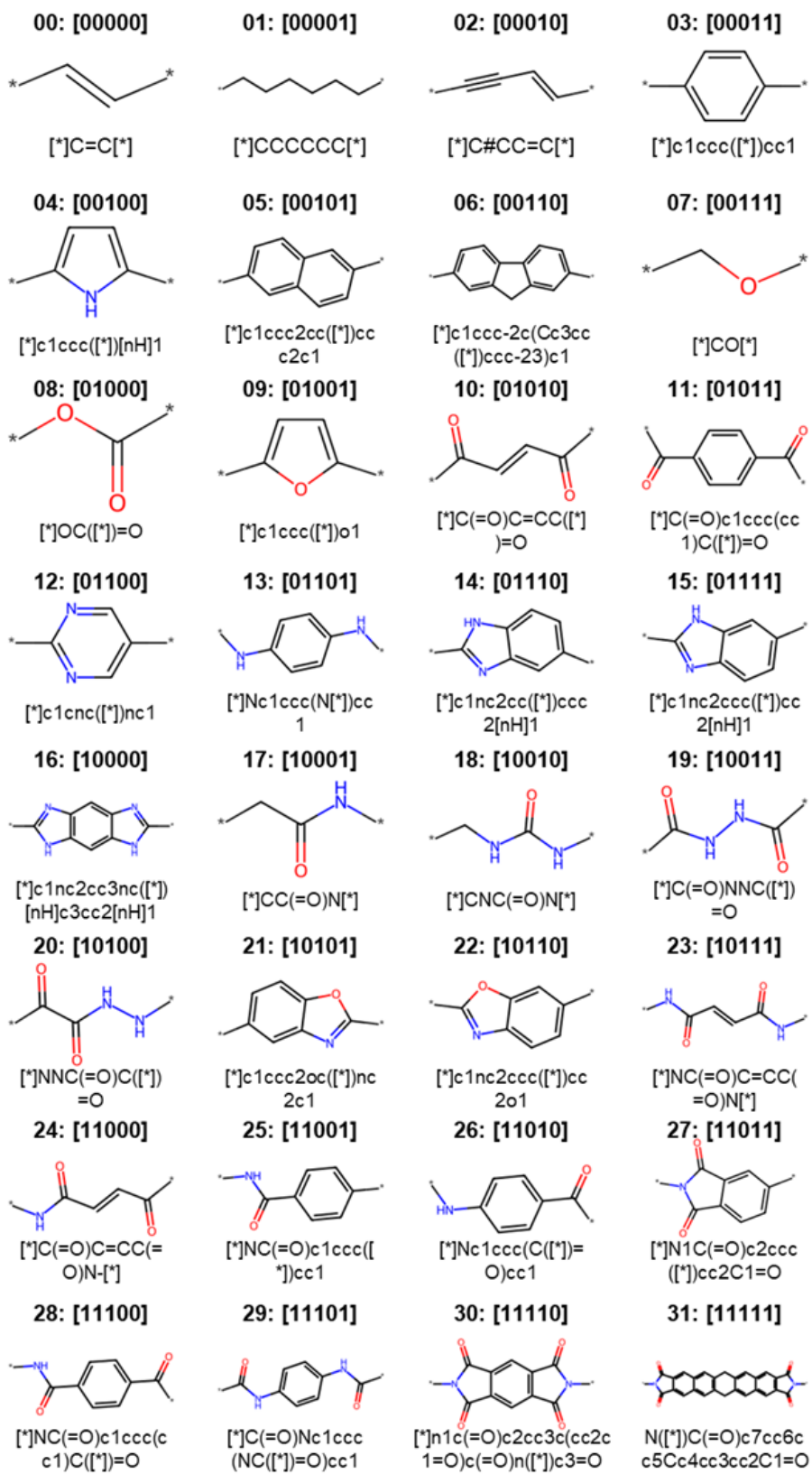

**Fig. S5.** Structures of 32 polymer fragments as basic units for high thermal conductivity polymer design.

## D. Convergence ability assessment of MOEA and MOBO algorithms

For obtaining statistical results we performed 20 runs of the MOEA and MOBO algorithms with different initial candidates, respectively, and the HV convergence curves are displayed in Figure S6a-b. Hypervolumes (HVs) of U-NSGA-III can rapidly rise to a certain level (within 20 generations), but it is difficult to increase again in subsequent. However, there are three qNEHVI runs that identified nine global optimal polymers within 200 generations and almost all of the HVs get a secondary boost after the first time to a certain level. The difference in this enhancement depends on the stochastic nature of QMC sampling. All the HVs of optimization algorithms reach a referred value that is calculated by the five ideal global optimal Pareto polymers and the referred point, although the mean HV of MOBO is greater than that of MOEA (see Figure S6c-d).

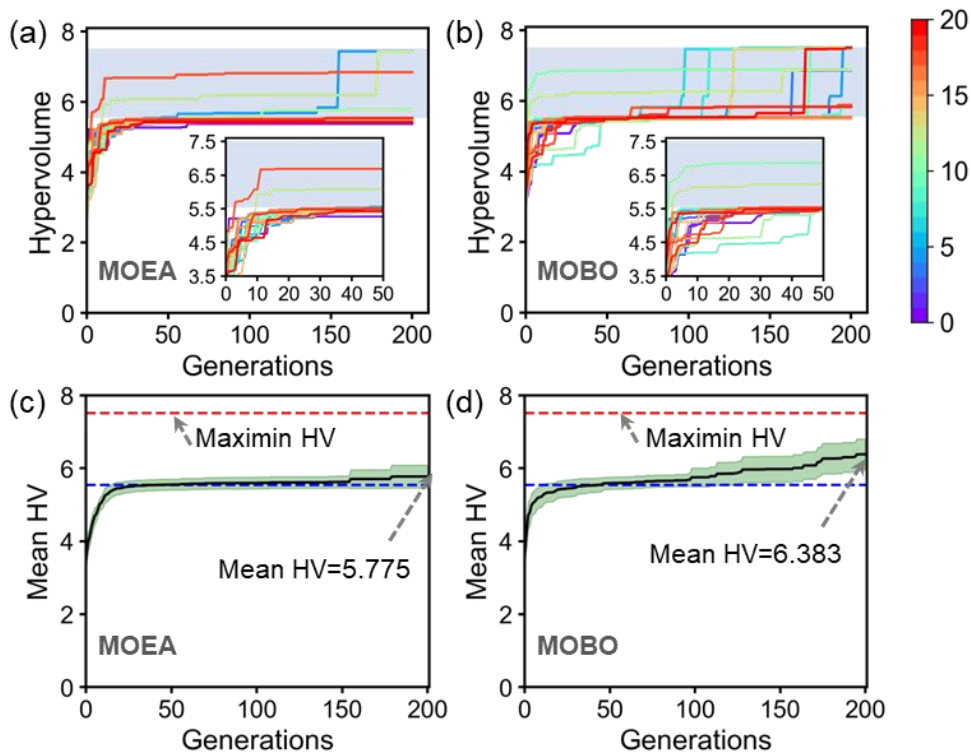

**Fig. S6.** Comparison of multi-objective evolutionary algorithm (MOEA) and multi-objective Bayesian optimization (MOBO) in triblock high thermal conductivity polymers inverse design. (a) and (b) Convergence curves for 20 runs of MOEA and MOBO. Each optimization run with 10 random initial structures and 200 iterations × 10 candidates per batch. (c) and (d) Mean hypervolume curves for 20 MOEA and MOBO runs. The upper edge of the blue strip or the red dashed line corresponds to the global optimal HV, and the lower edge of the blue strip or the blue dashed line indicates the HV computed from the five ideal global optimal Pareto polymers with the reference points

### E. Impact of initial structures on MOEA convergence performance

Starting from different initial structures, the convergence level of the HV curves of MOEAs varies widely, as shown in Fig. S7a. The red, green and blue lines correspond to 3 different sets of initial structures with random seeds of 12, 17 and 19, respectively. Each MOEA run has 10 initial structures and 10 candidates × 200 generations. Their optimization trajectories are displayed in Figure S7b~d. We noticed that various initial structures lead to differences in the optimization direction of the polymers. Among them, the initial structures controlled by seeds 12 and 17 are able to converge towards the entire Pareto front, while MOEA with seed 19 is only close to the four ideal Pareto polymers (TC > 0.40 W m$^{-1}$K$^{-1}$ and SA < 3.0). Therefore, MOEAs with seeds 12 and 17 have relatively large HVs (7.43 and 6.84) after 200 optimization iterations, while MOEA with seed 19 is only 5.43.

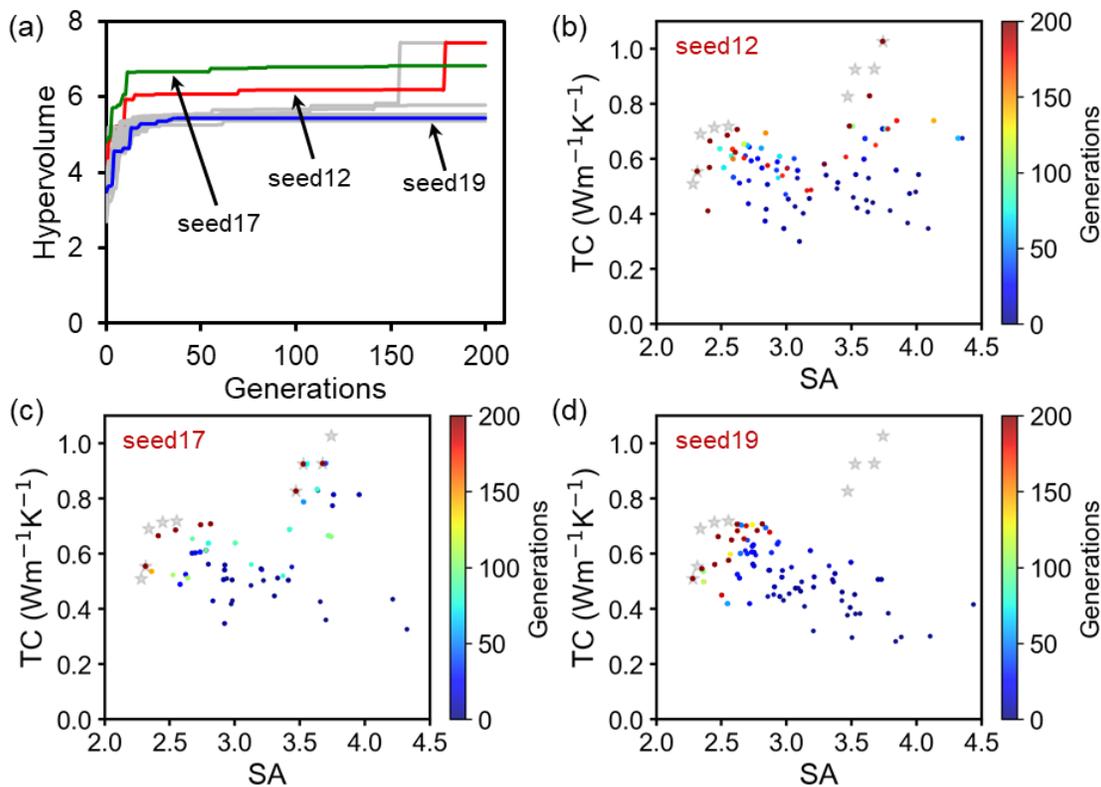

**Fig. S7.** Effect of initial structures on MOEA behavior. (a) HV convergence curves for MOEAs with different initial structures, where red, green, and blue lines correspond to MOEAs trained on the initial structures generated by the three random seeds of 12, 17 and 19, respectively. Their optimization trajectories are plotted in (b)~(d).

To demonstrate this point more intuitively, we counted the chemical blocks at different MOEA optimization stages, including 10 initial polymers with seed 12/17/19, and local Pareto polymers after 100 iterations and 200 iterations, as shown in Figure S8~S10. Moreover, chemical blocks of global Pareto polymers across the whole triblock polymers are presented for comparison. The 10 initial structures have diverse chemical blocks. Along with the optimization iterations, the types of chemical blocks of local Pareto polymers decrease and aggregate towards some promising blocks. In addition, chemical blocks in initial structures cover more blocks involved in the global Pareto polymers, which contributes to the convergence efficiency and capacity of MOEA.

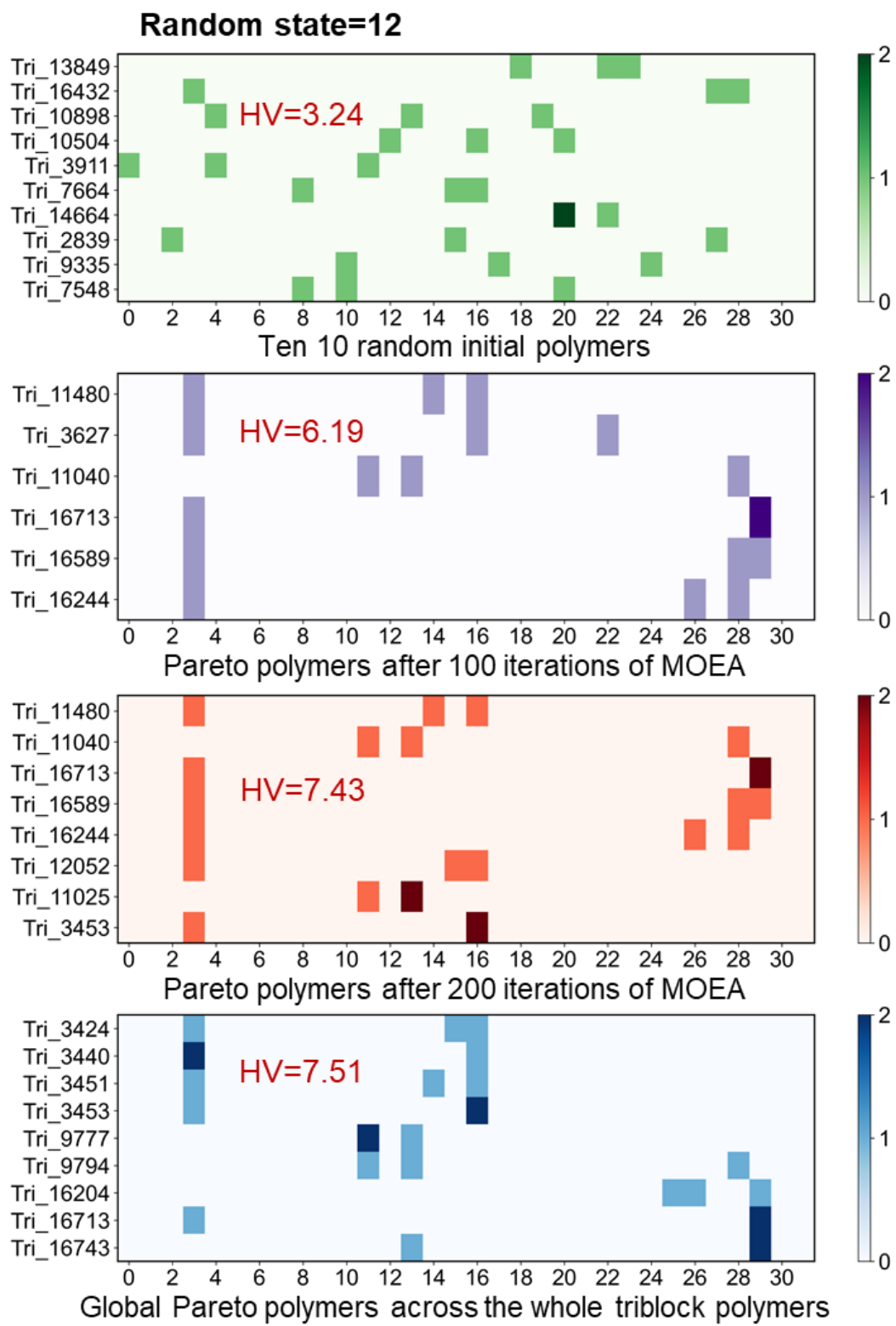

**Fig. S8.** Chemical blocks statistics for polymers at different stages, where 10 initial polymers were generated at the random seed of 12.

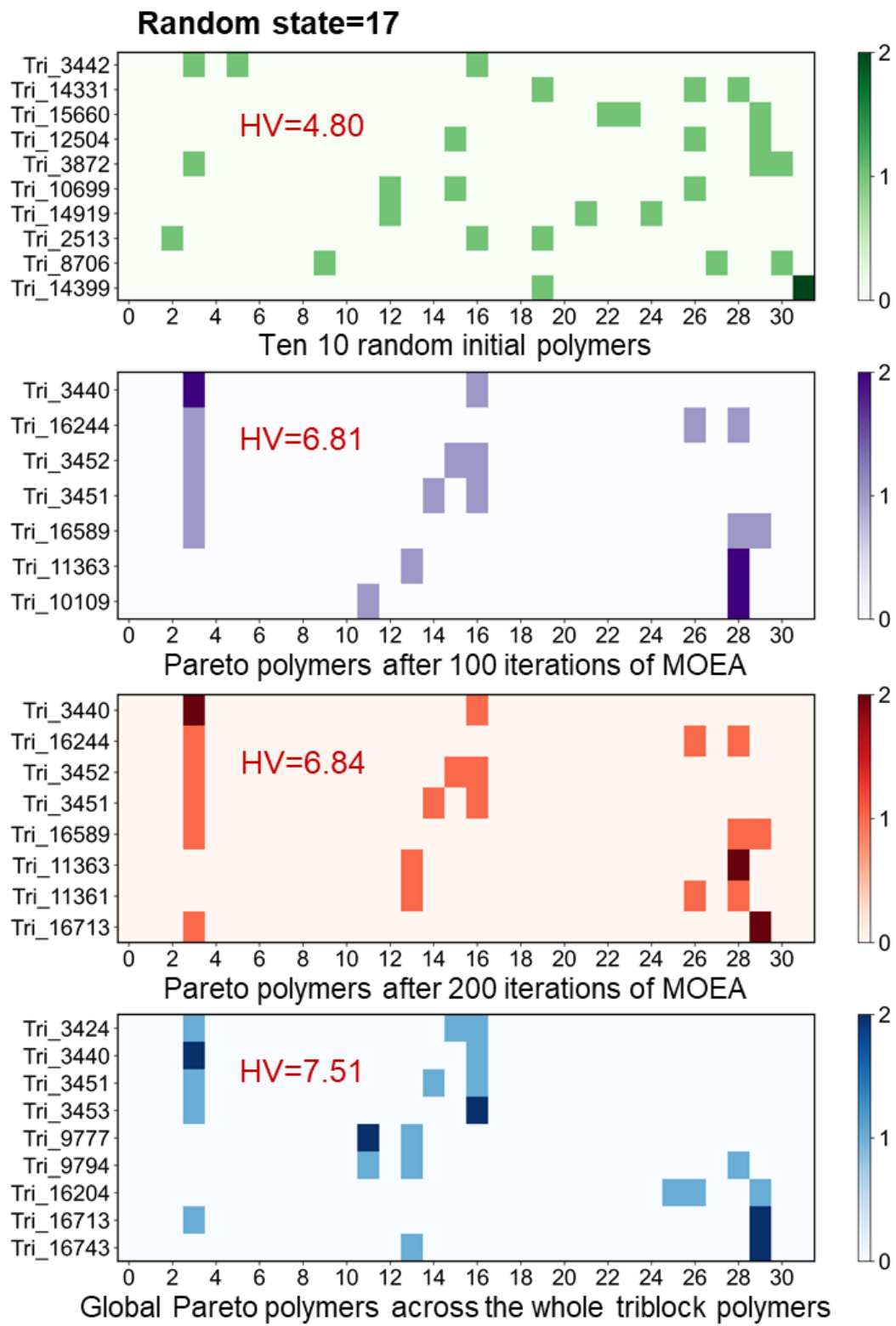

**Fig. S9.** Chemical blocks statistics for polymers at different stages, where 10 initial polymers were generated at the random seed of 17.

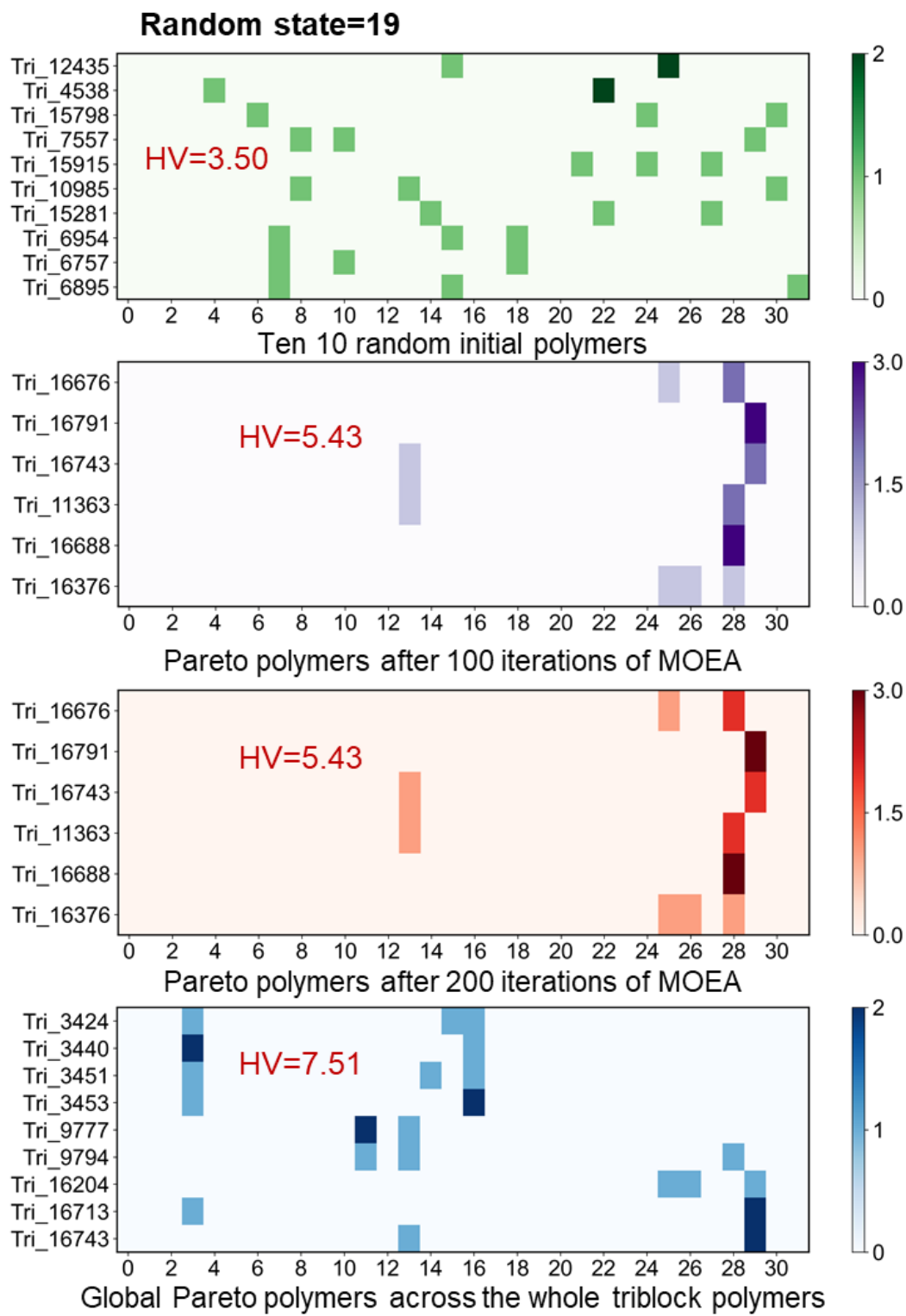

**Fig. S10.** Chemical blocks statistics for polymers at different stages, where 10 initial polymers were generated at the random seed of 19.

S13

## F. Parallel MOEAs versus MOBO for the design of high thermal conductivity pentablock polymers

Figure S11 demonstrates the distributions of generated pentablock polymers from 20 parallel MOEAs (blue dots) and a MOBO run (grey dots). Among 1921 MOEA-derived polymers, half of the candidates satisfy predefined requirements, i.e., SA≤3.0 and TC≥0.40 W m$^{-1}$K$^{-1}$. However, only 338 of 2005 polymers meet the above conditions in a MOBO run

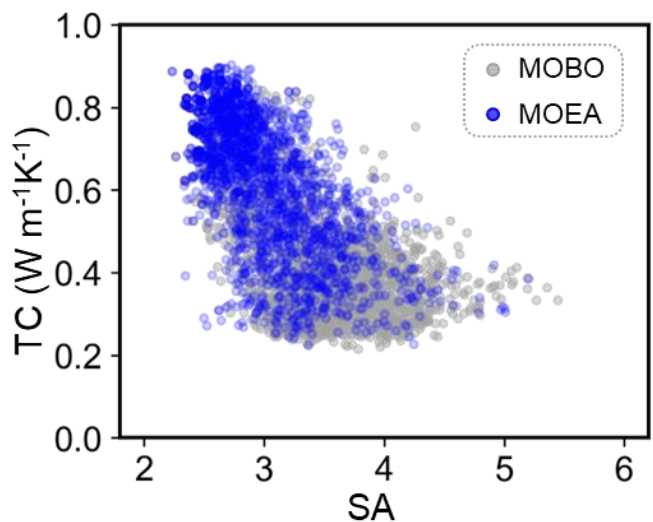

**Fig. S11.** Distributions of generated pentablock polymers from 20 parallel MOEAs (blue dots) and a MOBO run (grey dots).

**G. Demonstration of novel polymers with MD calculated TC in this work**

Table S2 lists the 50 promising polymers designed in this work, and their TCs were calculated by molecular dynamics simulation. The structures of repeating units can be viewed in the online tool of Marvin JS (https://marvinjs-demo.chemaxon.com/latest/ ) using SMILES as input.

**Table S2.** List of novel amorphous polymers (NAP) with MD calculated thermal conductivity, of which 20 triblock polymers and 30 pentablock polymers.

| No. | SMILES | TC (W m$^{-1}$K$^{-1}$) | Rg (Å) | SA score | Source |
|---|---|---|---|---|---|
| NAP_1 | [*]c1ccc(c2nc3cc4nc(c5cnc([*])nc5)[nH]c4cc3[nH]2)cc1 | 1.005 | 72.965 | 3.634 | Triblock polymer |
| NAP_2 | [*]c1ccc(c2nc3cc4nc(c5ccc([*])cc5)[nH]c4cc3[nH]2)cc1 | 0.777 | 76.485 | 3.471 | |
| NAP_3 | [*]c1ccc(c2nc3cc4nc(c5nc6cc7nc([*])[nH]c7cc6[nH]5)[nH]c4cc3[nH]2)cc1 | 0.753 | 74.992 | 3.741 | |
| NAP_4 | [*]c1ccc(c2nc3cc4nc(c5ccc6nc([*])oc6c5)[nH]c4cc3[nH]2)cc1 | 0.684 | 53.128 | 3.555 | |
| NAP_5 | [*]c1ccc(c2nc3cc4nc(c5ccc6nc([*])[nH]c6c5)[nH]c4cc3[nH]2)cc1 | 0.631 | 49.282 | 3.527 | |
| NAP_6 | [*]c1ccc(c2nc3cc(c4nc5cc6nc([*])[nH]c6cc5[nH]4)ccc3[nH]2)cc1 | 0.583 | 51.589 | 3.546 | |
| NAP_7 | [*]c1ccc(c2nc3cc4nc(c5nc6cc([*])ccc6o5)[nH]c4cc3[nH]2)cc1 | 0.571 | 54.191 | 3.699 | |
| NAP_8 | [*]c1ccc(c2nc3ccc(c4nc5cc6nc([*])[nH]c6cc5[nH]4)cc3[nH]2)cc1 | 0.570 | 50.309 | 3.527 | |
| NAP_9 | [*]c1ccc(c2ccc3nc(c4nc5cc6nc([*])[nH]c6cc5[nH]4)oc3c2)cc1 | 0.537 | 55.492 | 3.638 | |
| NAP_10 | O=C([*])Nc1ccc(NC(=O)Nc2ccc(NC(=O)Nc3ccc(NC(=O)[*])cc3)cc2)cc1 | 0.532 | 37.589 | 2.280 | |
| NAP_11 | [*]c1ccc(c2nc3cc4nc(c5nc6cc([*])ccc6[nH]5)[nH]c4cc3[nH]2)cc1 | 0.523 | 52.847 | 3.675 | |
| NAP_12 | O=C([*])Nc1ccc(NC(=O)c2ccc(C(=O)Nc3ccc(NC(=O)[*])cc3)cc2)cc1 | 0.494 | 50.556 | 2.314 | |
| NAP_13 | [*]c1ccc(c2ccc3oc(c4nc5cc6nc([*])[nH]c6cc5[nH]4)nc3c2)cc1 | 0.410 | 46.680 | 3.622 | |
| NAP_14 | O=C([*])c1ccc(C(=O)Nc2ccc(NC(=O)c3ccc(C(=O)N[*])cc3)cc2)cc1 | 0.408 | 29.009 | 2.443 | |
| NAP_15 | O=C([*])c1ccc(Nc2ccc(C(=O)Nc3ccc(C(=O)N[*])cc3)cc2)cc1 | 0.389 | 27.095 | 2.631 | |
| NAP_16 | O=C(Nc1ccc(NC(=O)Nc2ccc(C(=O)[*])cc2)cc1)c1ccc(C(=O)N[*])cc1 | 0.374 | 27.683 | 2.552 | |
| NAP_17 | O=C([*])c1ccc(NC(=O)c2ccc(Nc3ccc(C(=O)N[*])cc3)cc2)cc1 | 0.353 | 23.934 | 2.631 | |
| NAP_18 | O=C([*])c1ccc(NC(=O)c2ccc(C(=O)NC(=O)c3ccc(N[*])cc3)cc2)cc1 | 0.344 | 29.039 | 2.622 | |
| NAP_19 | O=C([*])c1ccc(C(=O)NC(=O)c2ccc(C(=O)NNc3ccc(N[*])cc3)cc2)cc1 | 0.321 | 27.350 | 2.814 | |
| NAP_20 | O=C([*])c1ccc(C(=O)Nc2ccc(NC(=O)c3ccc(C(=O)[*])cc3)cc2)cc1 | 0.315 | 26.576 | 2.336 | |

| No. | SMILES | TC (W m$^{-1}$K$^{-1}$) | Rg (Å) | SA score | Source |
|---|---|---|---|---|---|
| NAP_21 | O=C(NN[*])C(=O)c1ccc(c2ccc(C(=O)Nc3nc4ccc(c5nc6cc7nc([*])[nH]c7cc6[nH]5)cc4[nH]3)cc2)o1 | 0.904 | 32.468 | 3.647 | Pentablock polymer |
| NAP_22 | O=C([*])c1ccc(Nc2ccc(C(=O)NNC(=O)c3ccc(C(=O)OCC(=O)c4ccc(N[*])cc4)cc3)cc2)cc1 | 0.897 | 33.379 | 2.851 | |
| NAP_23 | O=C(Nc1ccc(NC(=O)OCC(=O)c2ccc(N[*])cc2)cc1)c1ccc(NC(=O)c2ccc(C(=O)[*])cc2)cc1 | 0.888 | 33.456 | 2.751 | |
| NAP_24 | O=C(O[*])Nc1ccc(NC(=O)c2ccc(C(=O)Nc3ccc(NC(=O)c4ccc(C(=O)[*])cc4)cc3)cc2)cc1 | 0.884 | 30.615 | 2.447 | |
| NAP_25 | O=C([*])c1ccc(C(=O)C=Cc2ccc(c3ccc4nc(c5ccc([*])cc5)[nH]c4c3)cc2)cc1 | 0.882 | 29.734 | 3.073 | |
| NAP_26 | O=C([*])Nc1ccc(NC(=O)C(=O)Nc2ccc(NC(=O)c3ccc(C(=O)NNC(=O)c4ccc(C(=O)Nc5ccc(C(=O)[*])cc5)cc4)cc3)cc2)cc1 | 0.880 | 28.939 | 2.685 | |
| NAP_27 | O=C(Nc1ccc(NC(=O)c2ccc(C(=O)N[*])cc2)cc1)c1ccc(C(=O)NNC(=O)c2ccc(C(=O)Nc3ccc(N[*])cc3)cc2)cc1 | 0.871 | 33.116 | 2.512 | |
| NAP_28 | O=C(Nc1ccc(NC(=O)c2ccc(C(=O)N[*])cc2)cc1)c1ccc(C(=O)NNC(=O)c2ccc(C(=O)Nc3ccc(C(=O)[*])cc3)cc2)cc1 | 0.862 | 40.238 | 2.538 | |
| NAP_29 | O=C(Nc1ccc(NC(=O)C(=O)Nc2ccc(NC(=O)c3ccc(C(=O)N[*])cc3)cc2)cc1)NC(=O)c1ccc(C(=O)Nc2ccc(C(=O)[*])cc2)cc1 | 0.840 | 27.541 | 2.783 | |
| NAP_30 | O=C(Nc1ccc(NC(=O)NC(=O)c2ccc(C(=O)Nc3ccc(NC(=O)c4ccc(C(=O)N[*])cc4)cc3)cc2)cc1)NC(=O)c1ccc(C(=O)[*])cc1 | 0.838 | 36.181 | 2.726 | |
| NAP_31 | O=C(c1ncc(Nc2ccc(Nc3cnc([*])nc3)cc2)cn1)c1ccc(NC=CC#C[*])cc1 | 0.801 | 32.844 | 3.666 | |
| NAP_32 | [*]c1ccc(CCCCCCc2ccc(CCCCCCCCCCCC[*])cc2)cc1 | 0.756 | 35.150 | 2.622 | |
| NAP_33 | [*]c1ccc(c2ccc3cc(c4ccc(c5ccc(c6ccc([*])cc6)cc5)cc4)ccc3c2)cc1 | 0.746 | 52.658 | 2.344 | |
| NAP_34 | O=C([*])c1ccc(NC(=O)c2ccc(C(=O)NNc3ccc(NC(=O)c4ccc(NC(=O)c5ccc(C(=O)N[*])cc5)cc4)cc3)cc2)cc1 | 0.715 | 32.926 | 2.627 | |
| NAP_35 | O=C(Nc1ccc(NC(=O)c2ccc(C(=O)Nc3ccc(NC(=O)c4ccc(C(=O)N[*])cc4)cc3)cc2)cc1)NC(=O)c1ccc(C(=O)[*])cc1 | 0.694 | 39.057 | 2.619 | |
| NAP_36 | O=C([*])c1ccc(C(=O)Nc2ccc(NC(=O)c3ccc(C(=O)Nc4ccc(NC(=O)c5ccc(C(=O)[*])cc5)cc4)cc3)cc2)cc1 | 0.680 | 29.537 | 2.234 | |
| NAP_37 | O=C(NNC(=O)C(=O)OC(=O)c1ccc(N[*])cc1)c1ccc(C(=O)c2ccc(C(=O)[*])cc2)cc1 | 0.676 | 28.847 | 3.099 | |
| NAP_38 | O=C([*])Oc1ccc(NC(=O)c2ccc(C(=O)Nc3ccc(NC(=O)c4ccc(C(=O)[*])cc4)cc3)cc2)cc1 | 0.669 | 46.822 | 2.462 | |

| No. | SMILES | TC (W m$^{-1}$K$^{-1}$) | Rg (Å) | SA score | Source |
|---|---|---|---|---|---|
| NAP_39 | O=C(C=CC(=O)c1ccc(NC(=O)c3nc4cc5nc(c6ccc7nc([*])oc7c6)[nH]c5cc4[nH]3)cc2)cc1)N[*] | 0.666 | 32.180 | 3.581 | |
| NAP_40 | O=C(Nc1ccc(NC(=O)OC[*])cc1)c1ccc(NC(=O)c2ccc(NC(=O)c3ccc(C(=O)[*])cc3)cc2)cc1 | 0.653 | 30.014 | 2.542 | |
| NAP_41 | O=C(Nc1nc2cc3nc(c4ccc(c5ccc([*])cc5)cc4)[nH]c3cc2[nH]1)c1ccc(C(=O)c2nc3ccc([*])cc3[nH]2)cc1 | 0.614 | 48.977 | 3.534 | |
| NAP_42 | O=C(Nc1nc2cc3nc(c4ccc(c5ccc([*])cc5)cc4)[nH]c3cc2[nH]1)c1ccc(C(=O)Nc2ccc(N[*])cc2)cc1 | 0.597 | 42.386 | 3.229 | |
| NAP_43 | O=C(Nc1ccc(NC(=O)c2ccc(C(=O)N[*])cc2)cc1)C(=O)Nc1ccc(NC(=O)C(=O)Nc2ccc(NC(=O)c3ccc([*])cc3)cc2)cc1 | 0.576 | 48.112 | 2.685 | |
| NAP_44 | O=C(Nc1ccc(NC(=O)C(=O)Nc2ccc(NC(=O)c3ccc(C(=O)N[*])cc3)cc2)cc1)Nc1ccc(NC(=O)c2ccc(C(=O)[*])cc2)cc1 | 0.537 | 29.749 | 2.656 | |
| NAP_45 | O=C(Nc1ccc(NC(=O)c2ccc(C(=O)Nc3ccc(C(=O)N[*])cc3)cc2)cc1)NC(=O)c1ccc(Nc2cc(C(=O)[*])cc2)cc1 | 0.520 | 24.420 | 2.702 | |
| NAP_46 | O=C([*])NNC(=O)c1ccc(C(=O)NNC(=O)c2ccc(C(=O)Nc3ccc(NC(=O)c4ccc(C(=O)N[*])cc4)cc3)cc2)cc1 | 0.465 | 24.866 | 2.571 | |
| NAP_47 | O=C([*])Nc1ccc(NC(=O)c2ccc(C(=O)Nc3ccc(NC(=O)c4ccc5cc(C(=O)Nc6ccc(NC(=O)[*])cc6)ccc5c4)cc3)cc2)cc1 | 0.460 | 24.916 | 2.543 | |
| NAP_48 | O=C(Nc1ccc(NC(=O)c2ccc(C(=O)O[*])cc2)cc1)c1ccc(NC(=O)c2ccc(C(=O)[*])cc2)cc1 | 0.395 | 28.573 | 2.425 | |
| NAP_49 | O=C(Nc1ccc(NC(=O)c2ccc(C(=O)Nc3ccc(NC(=O)c4ccc(C(=O)N[*])cc4)cc3)cc2)cc1)c1ccc([*])cc1 | 0.393 | 74.187 | 2.416 | |
| NAP_50 | O=C(Nc1ccc(NC(=O)c2ccc(C(=O)Nc3ccc(C(=O)N[*])cc3)cc2)cc1)c1ccc(NC(=O)c2ccc(C(=O)[*])cc2)cc1 | 0.309 | 21.521 | 2.367 | |

## H. Expansion case of parallel MOEAs for designing triblock polymers with TC > 0.40 Wm$^{-1}$K$^{-1}$ and RI > 1.80

Our proposed scheme is flexible and universal, and can be extended to the design of polymers with other target properties. To demonstrate it, we provide an expansion case of parallel MOEAs for designing triblock polymers with TC > 0.40 Wm$^{-1}$K$^{-1}$ and RI (refractive index) > 1.80. Polymers with high RI are favorable for flexible displays [5], organic light-emitting diodes [6] and image sensors [7]. Moreover, the computational database records the RI of polymers, which was calculated by Psi4 using Lorentz–Lorenz equation in RadonPy [8,9]. There are 1138 candidates with known RIs out of 1144 polymers in the benchmark dataset. Based on these 1138 polymers, we trained a random forest (RF) model in Scikit-learn using five-fold cross-validation and Bayesian optimization for the determination of hyperparameters [2]. Figure S12 displays the pairs of RI predicted by the RF model versus that calculated by the MD simulation, which suggests that the trained surrogate model has good predictive ability, with R$^2$ of 0.93±0.02 and RMSE of 0024±0.03.

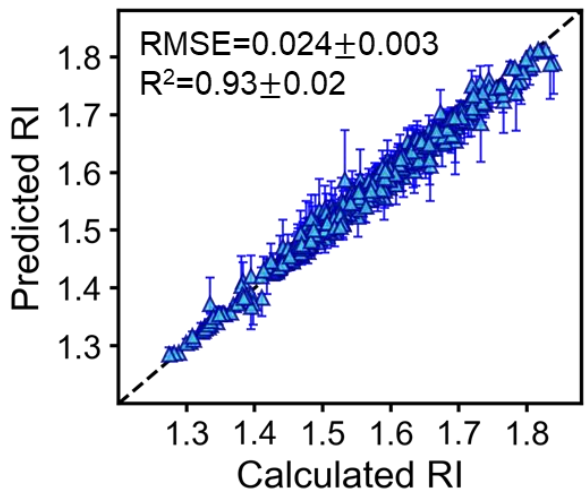

**Fig. S12.** Random forest model for refractive index prediction

The entire 16896 triblock polymers were configured to the exploration space, and their RIs versus TCs are shown in Figure S13a. There are eight polymers at the global Pareto front, and 7/8 are polyimines, as illustrated in Figure S13d. Our optimization target was set to TC > 0.40 W m$^{-1}$K$^{-1}$ and RI > 1.80. However, achieving a high refractive index (> 1.80) is not easy, with a percentage of only 0.38% among all triblock polymers (Figure S13b). The Venn diagram of Figure S13c counts the number of target polymers in the whole exploration space as 56, with a ratio of only 0.33%.

We performed 10 MOEAs in parallel for designing triblock polymers with target properties. Each MOEA starts from 10 randomized initial structures, and goes through 10 candidates × 200 generations. Figure S14a exhibits the HV curves of 10 MOEA runs, where the reference point was set to [0.15,1.45] for TC and RI, and the maximum HV value is 3.83 (Calculated from 8 global Pareto polymers and the reference point). The final HVs ranged from 2.80 to 3.50, since different collections of initial candidates. We then employed the Gaussian kernel to estimate the probability density function (PDF) of all searched polymers in 10 MOEAs, as shown in Figure S14b. The high probability region occurs close to the global Pareto front, which reflects the robustness of the parallel MOEA. Figure S14c illustrates 91 unrepeated triblock polymers (blue dots) designed by parallel MOEAs, of which 16 candidates (16.7%, in Figure S14d) are the target polymers. This demonstrates the scalability of our developed inverse design workflow.

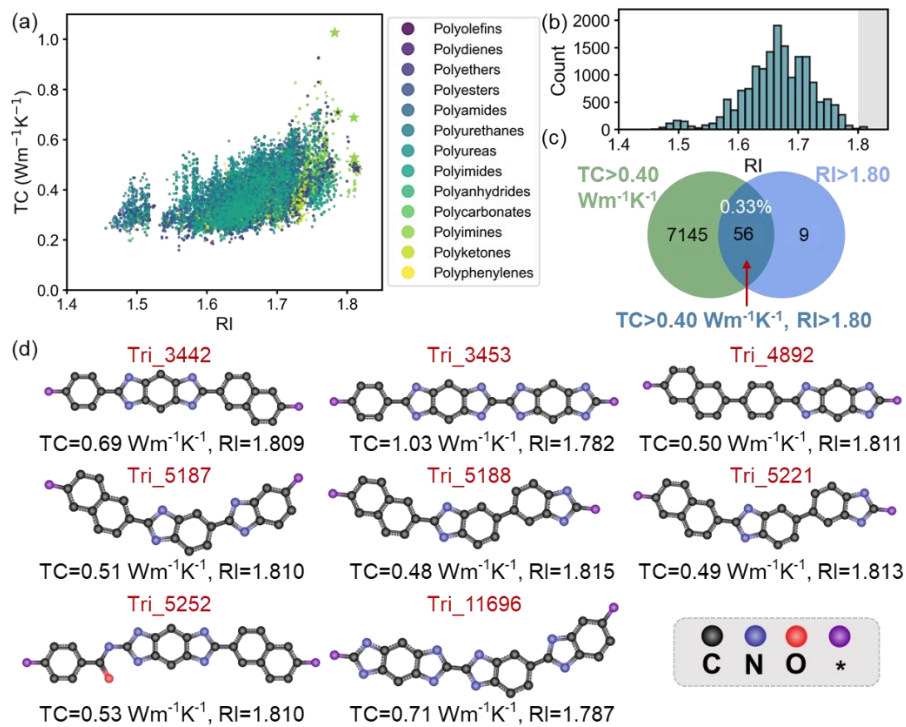

**Fig. S13.** All triblock polymers with ML predicted TCs and RIs. (a) Relationship between RIs and TCs for 16896 triblock polymers. (b) Distribution of RIs. (c) Venn diagram statistics for polymers with TC>0.40 Wm$^{-1}$K$^{-1}$ or RI>1.80. (d) Repeating units of eight global Pareto polymers.

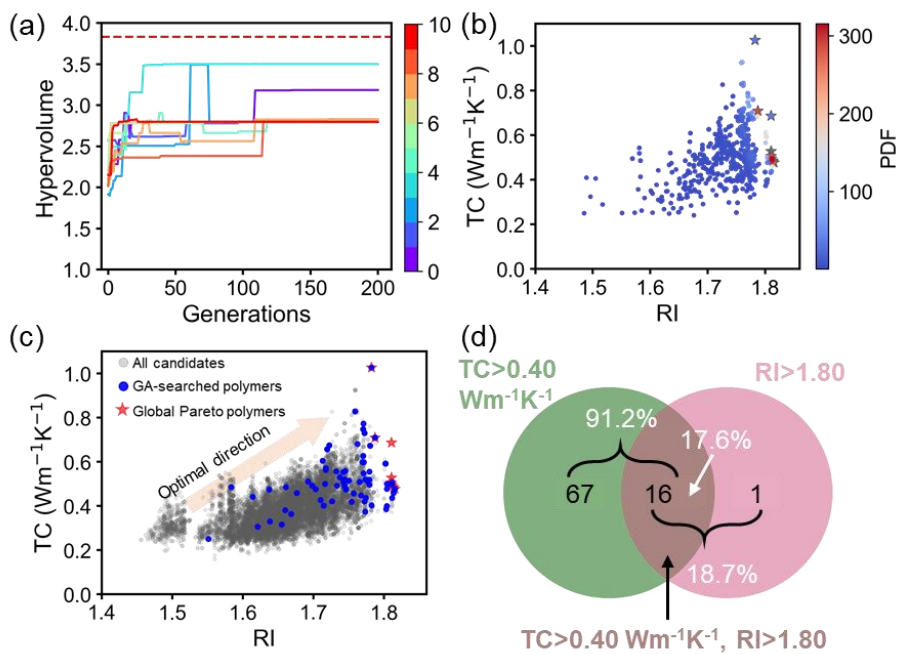

**Fig. S14.** Parallel MOEAs for the design of innovative triblock polymers. (a) HV curves for 10 MOEAs with different initial structures. (b) Probability density maps in objective space for 10 runs of MOEA. (c) Distributions of 91 non-repeating triblock polymers generated by 10 parallel MOEAs. (d) Venn diagram statistics for MOEA-produced polymers with TC > 0.40 Wm$^{-1}$K$^{-1}$ or RI > 1.80.